\documentclass[prb,twocolumn,showpacs]{revtex4}

\usepackage{graphicx}
\usepackage{dcolumn}
\usepackage{amsmath}
\usepackage{amssymb}
\usepackage{xspace}
\usepackage{ifthen}

\newlength{\ldag}
\newcommand{\ie}{i.e.\@\xspace}

\newcommand{\adag}{{a^\dagger}}
\newcommand{\ana}{{a^{\phantom\dagger}\hspace{-\ldag}}}
\newcommand{\emku}{\eta_{\mbox{\tiny{MKU}}}}
\newcommand{\ew}{\eta_{\mbox{\tiny{W}}}}
\newcommand{\rmc}{\mathrm c}
\newcommand{\rmd}{\mathrm d}
\newcommand{\rmi}{\mathrm i}
\newcommand{\rmr}{\mathrm R}
\newcommand{\rmb}{\mathrm B}
\newcommand{\pal}{\partial_l}

\newcommand{\ve}{\varepsilon}

\newcommand{\compare}[3]{
\ifthenelse{\lengthtest{#1 > #2}}{\setlength{#3}{#1}}
{\setlength{#3}{#2}}}
\newlength{\ecart}
\newlength{\qqch}
\newlength{\longueura}
\newlength{\longueurb}
\newlength{\longueurc}
\newlength{\longueurd}
\newlength{\maxab}
\newlength{\maxcd}
\newlength{\longueure}
\newlength{\longueurf}
\newlength{\maxef}
\newcommand{\xxx}[6]{
\settowidth{\longueura}{\scriptsize $#1$}
\settowidth{\longueurb}{\scriptsize $#2$}
\settowidth{\longueurc}{\scriptsize $#3$}
\settowidth{\longueurd}{\scriptsize $#4$}
\settowidth{\longueure}{\scriptsize $#5$}
\settowidth{\longueurf}{\scriptsize $#6$}
\compare{\longueura}{\longueurb}{\maxab}
\compare{\longueurc}{\longueurd}{\maxcd}
\compare{\longueure}{\longueurf}{\maxef}
\addtolength{\maxab}{\qqch}
\addtolength{\maxcd}{\qqch}
\addtolength{\maxef}{\qqch}
{}_{\begin{minipage}[b]{\maxab}
\centering \scriptsize $#1$ \end{minipage}
\hspace*{\ecart},\hspace*{\ecart}
\begin{minipage}[b]{\maxcd}
\centering \scriptsize $#3$ \end{minipage}
\hspace*{\ecart},\hspace*{\ecart}
\begin{minipage}[b]{\maxef}
\centering \scriptsize $#5$ \end{minipage}}
^{\begin{minipage}[b]{\maxab}
\centering \scriptsize $#2$ \end{minipage}
\hspace*{\ecart},\hspace*{\ecart}
\begin{minipage}[b]{\maxcd}
\centering \scriptsize $#4$ \end{minipage}
\hspace*{\ecart},\hspace*{\ecart}
\begin{minipage}[b]{\maxef}
\centering \scriptsize $#6$ \end{minipage}}
}

\begin{document}
\graphicspath{{/EPSF/}{figures/}{./}} 

\title{The Quartic Oscillator: 
a Non-Perturbative Study by 
Continuous Unitary Transformations}

\author{S\'ebastien Dusuel$^{1,3}$, G\"otz S. Uhrig$^{2,1}$}

\affiliation{
$^1$Institut f\"ur Theoretische Physik, Universit\"at zu K\"oln,
Z\"ulpicher Str. 77, 50937 K\"oln, Germany\\
$^2$Department of Physics, Tohoku University, Sendai 980-8578, Japan\\
$^3$Laboratoire de Physique, 
UMR-CNRS 5672, ENS Lyon, 46 All\'ee d'Italie, 69364 Lyon Cedex 07, France}


\begin{abstract}
The quantum quartic oscillator is investigated in order to test the
many-body technique of the continuous unitary transformations. 
The quartic oscillator is sufficiently simple to allow a detailed study and
comparison of various approximation schemes.  Due to its simplicity, it can 
be used as pedagogical introduction in the unitary transformations. 
Both the spectrum and the spectral weights are discussed.
\end{abstract}


\pacs{11.10.Gh,11.10.Hi,05.10.Cc,03.65.Db}

\maketitle

\section{Introduction}
\label{sec:intro}

Quantum many-body systems often can be simplified considerably by choosing
an appropriate basis. This is the first step where the qualitative
understanding of the system enters.  If the problem is first formulated
in a basis which is not optimum for the subsequent treatment it is
reasonable to change the basis. This is a unitary transformation. 
Famous examples of such changes of basis are the fermionic one that leads
to the  wave function of the superconducting BCS phase or the bosonic
Bogoliubov transformation which leads to the spin waves in a Heisenberg quantum
antiferromagnet. These examples refer to mean-field treatments. 
The complete solution of a many-body problem by a simple analytical
transformation is rarely possible. 

Ten  years ago, the proposal was made to use unitary transformations
beyond the single particle level \cite{wegne94,glaze93,glaze94}. The
unitary transformation is performed in a continuous fashion
whence the name continuous unitary transformation (CUT). This
means that it is parametrized by a continuous parameter $l$ running from
0 to $\infty$. At $l=0$ the Hamiltonian is unchanged and at $l=\infty$
it is in its final form. The fundamental idea is that the transformation
at a later stage, \ie, at larger values  of $l$, can take into account
the changes introduced already at an earlier stage. The transformation
can be set up such that processes at larger energy are treated before
those at lower energies. Then the transformation has properties
similar to Wilson's renormalization group approach \cite{wilso75}.
The objective is of course to simplify the model by the transformation.
At best, the final Hamiltonian has become diagonal.

The aim of the present paper is to illustrate the CUT approach,
and to study it for a simple model where exact (numerical) data is also
easily accessible. The simplicity of the model permits to present the
equations on the basic level and also to go to higher orders. Our 
presentation is intended to serve as introduction to the CUT method. This will
enable the interested reader to generalize the approach to more elaborate 
models.

The  model of our choice is the quantum anharmonic oscillator which is 
defined by the usual quadratic kinetic energy and
a potential with quadratic and quartic parts. We will abbreviate
the oscillator comprising both a quadratic and a quartic part in the
potential by QO; the one which has only a quartic one is called the
pure quartic oscillator (PQO).
This model has often been investigated as testbed for more complicated
situation, e.g., it was used to test other renormalization treatments
\cite{aoki02,hedde04}.

The bottom line of our investigation is that the CUT method works extremely 
well providing very good quantitative results. 
The starting point of the CUTs can easily be optimized by incorporating  
prior understanding of the physics of the underlying model.
Then only very moderate effort suffices to obtain satisfactory results.

Following this Introduction, we present basic facts and insights 
about the CUTs. In Sect.\ III results for the energetically
lowest states of the QO and the PQO will be shown for various ways to conduct 
the transformation. Spectral weights will 
also be computed in this section.
In Sect.\ IV we discuss in which way the CUT approach can be systematically 
improved. 
Sect.\ V concludes the article.


\section{Basics about the  CUTs}
\label{sec:learning}

Before addressing the anharmonic oscillator, it is worthwhile
to study the harmonic oscillator in order to become familiar 
with the CUTs.

\subsection{Method}
A given Hamiltonian $H$ can be diagonalized by a 
suitable unitary transform $U$ 
($U U^\dagger=\mathbb{I}$). But it is usually a very hard task to find 
this transform $U$. The idea of Wegner\cite{wegne94} (and independently of 
G{\l}azek and  Wilson\cite{glaze93,glaze94}) was to diagonalize the 
Hamiltonian  in a continuous way starting from the original Hamiltonian 
$H=H(0)$. A flowing Hamiltonian is defined by  $H(l)=U^\dagger(l) H U(l)$ 
depending on  
the parameter $l$  such that $H(l=\infty)$ is decisively simpler. 

The continuous transform  is equivalent to performing  infinitely many 
infinitesimal unitary transforms $e^{-\eta(l) \rmd l}$ with the
generator  $\eta(l)=-U^\dagger(l)\pal U(l)$. The flow equation
\begin{equation}
  \label{eq:dlH}
  \pal H(l) = [\eta(l),H(l)]
\end{equation}
defines the change of the Hamiltonian. Of course, the whole transformation
$U(\infty)=\mathcal{L} \exp(-\int_0^\infty \eta(l) \rmd l)$ can also be given 
in one formal expression. Note, however, that this requires the use of the
$l$-ordering operator $\mathcal{L}$ which sorts from left to right
according to ascending values of $l$. This means that the whole transformation
operator can be extremely complicated. For simple models, however, the explicit
transformation can also be written down.

The crucial point is to choose the generator $\eta$ which leads to 
a simplification of the Hamiltonian. Wegner proposed to take the commutator 
between the  diagonal part $H_\rmd$ and the non-diagonal part $H_\mathrm{nd}$
in a given basis. This means that the generator reads
$\ew(l)=[H_\rmd(l),H_\mathrm{nd}(l)]=[H_\rmd(l),H(l)]$. 
In the  basis of the eigen states of $H_\rmd$ the generator
thus reads 
\begin{equation}
  \label{eq:wegner_generaor_general}
\eta_{i,j}(l)=\Big(\ve_i(l)-\ve_j(l)\Big)H_{i,j}(l)\ .
\end{equation}
For finite matrices, this choice was  proven \cite{wegne94} to achieve 
\begin{equation}
  [H_\rmd(\infty),H(\infty)]=0 \ .
\end{equation}
If $H_\rmd(\infty)$ is non-degenerate this equation implies that
$H(\infty)$ is diagonal. Otherwise it implies block-diagonality with
respects to the degenerate subspaces of $H_\rmd(\infty)$.
An extension of this proof to infinite matrices is given in Appendix 
\ref{app:proof}. Note that there is an enormous freedom
 in what one considers to be diagonal and non-diagonal.  For 
practical calculations the only restriction on $H_\rmd$ should be that it can
easily be diagonalized.

For Hamiltonians of the form of band matrices, Mielke proposed 
another generator involving the sign of the difference of the vector indices 
(see below). This generator preserves the band structure during the flow
\cite{mielk98} which  is not the case for Wegner's generator.
Independently, Knetter and Uhrig observed that a generator based
on the sign of the difference of the particle number allows
to construct particle-number conserving  effective Hamiltonians
\cite{uhrig98c,knett00a}. 
If $Q$ is the  operator counting the number of  elementary excitations 
and the matrix elements of $\eta$ in the eigen basis of $Q$ are chosen to be
\begin{equation}
  \label{eq:mku_generaor_general}
  \eta_{i,j}(l) = \mathrm{sgn}\Big(q_i(l)-q_j(l)\Big)H_{i,j}(l)
\end{equation}
and the final Hamiltonian satisfies $[Q,H(\infty)]=0$. Furthermore,
one can show that this type of generator sorts the eigen values in 
ascending order of the particle number of the corresponding eigen vectors
\cite{mielk98,knett00a}.
Analogous generators were also used by Stein \cite{stein97,stein98} for
models where the sign-function is not required since only one finite
absolute value of $\Delta q := q_i-q_j$ occurs. For instance, if
$\Delta q = 0, \pm2$ the sign-function only changes the scale of
$l$. But for $\Delta q = 0, \pm2,\pm4$ the sign is essential.

In the sequel, the generators of Wegner's type will be denoted by the subscript
W; the generators characterized by the sign as introduced
by Mielke, Knetter and Uhrig will be denoted by the 
subscript MKU. Any coefficient $g$ without argument
appearing in the Hamiltonian or in the generator stands for the running 
coefficient $g(l)$. Their initial values are denoted by $g^{(\rmb)}$
(bare) and their final values by $g^{(\rmr)}$ (renormalized).


\subsection{Two Solvable Models}
\label{sec:sub:gaussian_models}

We illustrate the above for easily solvable models, namely a two-level 
system and a harmonic oscillator. The Hamiltonians read
\begin{subequations}
  \begin{eqnarray}
    \label{eq:ham_ferm_quad}
    H_2&=&E\mathbb{I} - \frac{\omega}{2} \sigma^z + \frac{e}{2}\sigma^x,\\
    \label{eq:ham_bos_quad}
    H_\mathrm{HO} &=& E\mathbb{I} + \omega\adag \ana + \frac{d}{2}\left( \adag^2+
      \ana^2\right) \ ,
  \end{eqnarray}
\end{subequations}
where $\sigma^i$ are Pauli matrices, and $ a$ and $\adag$ are bosonic 
annihilation and creation operators satisfying the commutation relation
$[ a,\adag]=1$.

\begin{figure}[htbp]
  \centering
  \includegraphics[width=4cm]{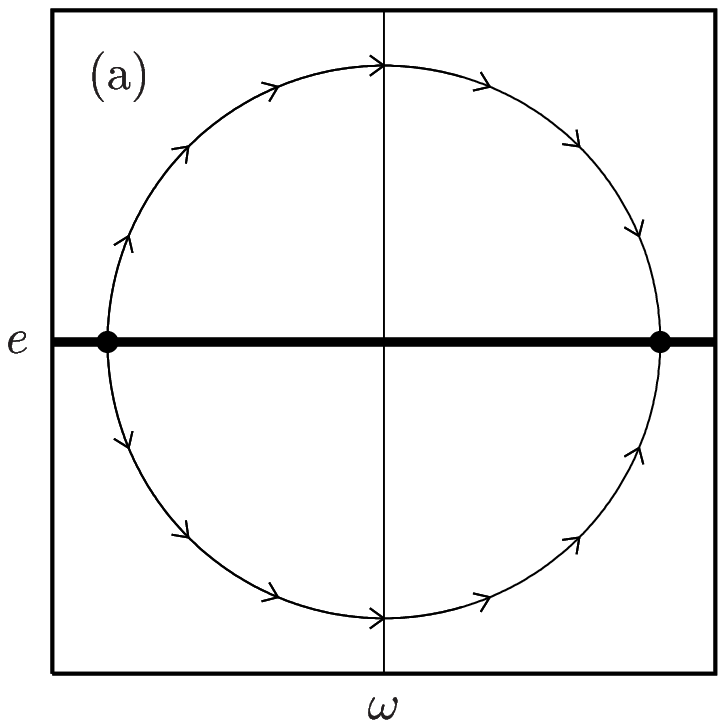}
  \hspace{0.2cm}
  \includegraphics[width=4cm]{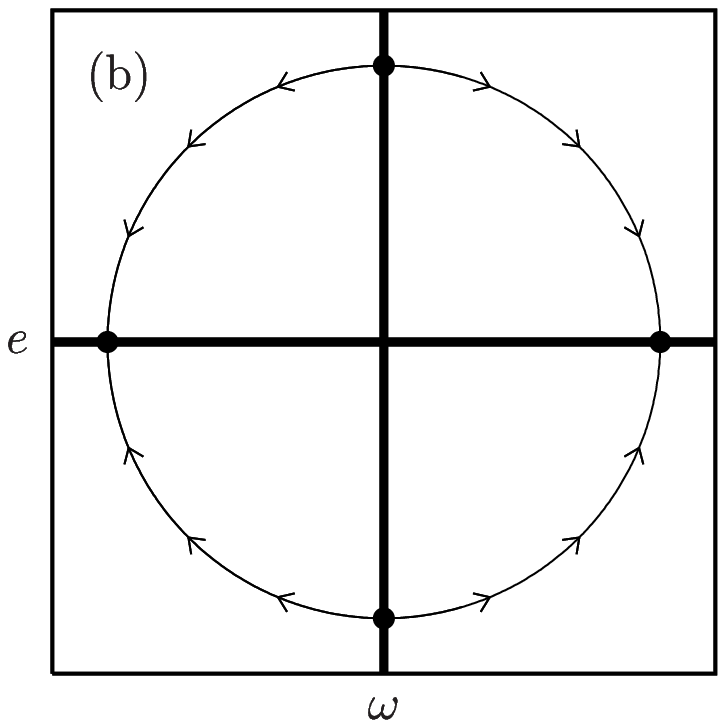} 
  \caption{Flows for the two-level system. Panel (a) depicts the flow
from the MKU-generator; Panel (b) the one from Wegner's generator. Thick lines 
represent lines of fixed points; the dots represent the fixed points of
 the particular circular trajectory shown. The arrows indicate the 
direction of the flows for increasing $l$.}
  \label{fig:OQ_fermion}
\end{figure}
Let us first consider the two-level system. The counting operator  is 
$Q=(\mathbb{I}-\sigma^z)/2$. Hence, the MKU-generator is given by 
$\emku=-(e/2)\rmi\sigma^y$. The flow equations read
\begin{subequations}
  \begin{eqnarray}
    \label{eq:flow_QO_fermion_mku}
    \pal E &=&0\\
    \pal \omega&=&e^2\\
    \pal e&=&-\omega e\ .
  \end{eqnarray}
\end{subequations}

The diagonal part of the Hamiltonian is the one proportional to $\omega$.
Thus Wegner's generator is found to equal  
$\ew=[H_\rmd,H_\mathrm{nd}]=[-(\omega/2)\sigma^z,(e/2)\sigma^x]=\omega\emku$. 
So the flow equations are the same as above if we rescale the continuous
parameter $l \to l'$ with $\rmd l'=\omega(l) \rmd l$ and $l'(l=0)=0$. 
The flow trajectories in $(\omega,e)$ space are circles since 
$\omega^2+e^2=R^2$ is a constant of the flow, see Fig.~\ref{fig:OQ_fermion}.
Thus the diagonalized Hamiltonian has the coefficients 
$\omega^{(\rmr)}=\sqrt{R}=\sqrt{{\omega^{(\rmb)}}^2+{e^{(\rmb)}}^2}$ and 
$E^{(\rmr)}=E^{(\rmb)}$.
\begin{figure}[htbp]
  \centering
  \includegraphics[width=4cm]{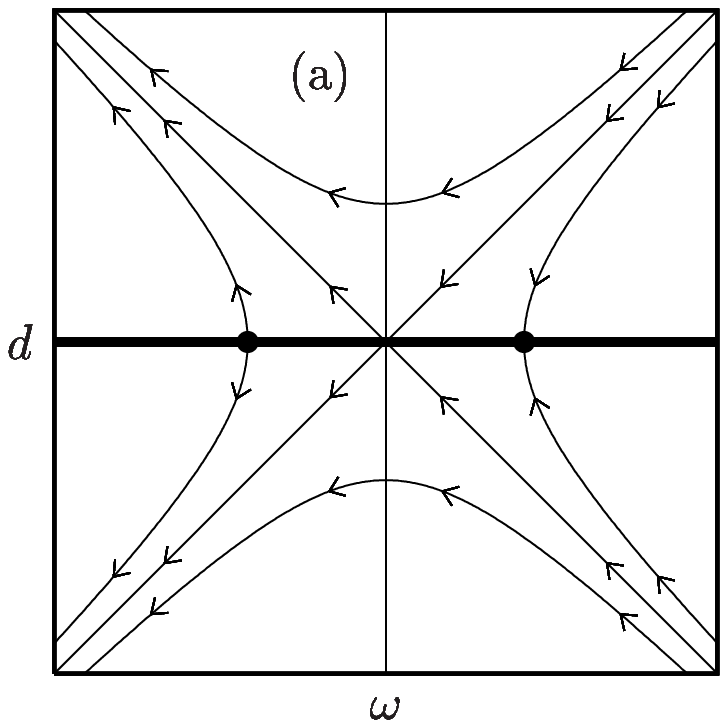}
  \hspace{0.2cm}
  \includegraphics[width=4cm]{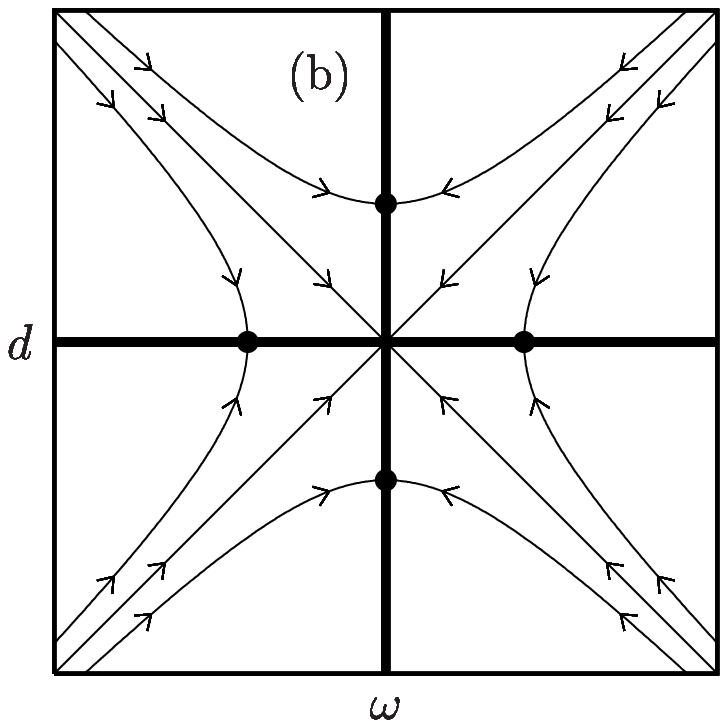}
  \caption{Flows for the bosonic oscillator. (a) MKU-generator and (b) Wegner's
    generator. Thick lines represent lines of fixed points, and the dots 
    represent the fixed point reached for the particular hyperbolic trajectory 
    shown. The arrows indicate the direction of the flows for growing $l$.}
  \label{fig:OQ_boson}
\end{figure}

Second, we consider the bosonic oscillator for which the counting operator
is $Q=\adag \ana$. So the MKU-generator is $\emku=d/2(\adag^2- \ana^2)$ 
and the flow equations read
\begin{subequations}
  \begin{eqnarray}
    \label{eq:flow_OQ_boson}
    \pal E &=& -d^2 \\
    \pal \omega &=& -2d^2 \\
    \pal d & =& -2\omega d \ .
  \end{eqnarray}
\end{subequations}
The canonical choice of the diagonal part is $H_\rmd = \omega\adag \ana$.
Thus we have again $\ew=[\omega\adag \ana,d/2(\adag^2+ \ana^2)]=\omega\emku$
so that the flow trajectories of both generators are the same.
The expression $\omega^2-d^2$ is a constant of the flow so that the 
trajectories are hyperbolae as shown in  Fig.~\ref{fig:OQ_boson}. 
For a hermitian Hamiltonian bounded below, \ie, $\omega>|d|$ (see below), 
the final coefficients of the diagonalized Hamiltonian read 
$\omega^{(\rmr)}=\sqrt{{\omega^{(\rmb)}}^2-{d^{(\rmb)}}^2}$ and 
$E^{(\rmr)}=E^{(\rmb)}+(\omega^{(\rmb)}-\omega^{(\rmr)})/2$. The
energy is found by integration of $\pal E=\pal \omega/2$.

As expected from the rescaling $\rmd l\to \rmd l'=\omega\rmd l$
the topologies of both flows are the same. The directions are the same
for positive values of $\omega$.
Only for negative values of $\omega$ the directions differ.
The coincidence of the topologies of the flows for both choices of
the generator is special to the simple systems under study here.
It is not a general feature for more complicated, generic  Hamiltonians.


\subsection{Remarks}
\label{sec:sub:insights}

First, we want to stress that the CUTs exactly diagonalize these simple 
Hamiltonians. Note that for the harmonic oscillator, the CUTs are infinitesimal
 Bogoliubov transformations  whose total effect corresponds to 
 the single-step Bogoliubov  transformation diagonalizing the 
Hamiltonian.\cite{ohira02} This was to be expected since there is only
one unique way, except for phases, to diagonalize non-degenerate problems.

Second, we see that Wegner's generator always leads to a fixed point.
This does not need to correspond to a final diagonal Hamiltonian when
there are degeneracies as it is the case for $\omega=0$. In contrast
the MKU-generator is not sensitive to degeneracies, but it does not
always lead to a fixed point. A fixed direction may occur
as seen in Fig.~\ref{fig:OQ_boson}(a).\footnote{See Ref.\ \onlinecite{dusue02}
for a detailed discussion of fixed directions.}
The flow in Fig.~\ref{fig:OQ_boson}(a) is similar to the Kosterlitz-Thouless 
flow. The fixed directions are reached when the Hamiltonian is unbounded 
below, \ie $\omega<|d|$. In this case the proof of convergence indeed fails.
\footnote{The proofs of convergence that 
can be found for the MKU-generator\cite{mielk98,knett00a} and for Wegner's 
generator\cite{wegne94} are valid for finite matrices. In Appendix 
\ref{app:proof}, the proof of convergence for Wegner's generator is extended 
to infinite matrices.} A possible cure to this problem is 
to diagonalize $-H$, or what amounts to the same, to revert the flows.
\cite{ohira02} Physically, Hamiltonians unbounded below
occur only for open physical systems whose treatment requires special care.
This is beyond the scope of the present investigation.
It is a general feature that Wegner's generator always leads to a fixed point 
and that it is stopped by degeneracies. In contrast the MKU generator is 
insensitive to degeneracies, but it may fail to converge to a fixed point.


\section{The Quartic Oscillator}
\label{sec:QO_lowest_order}

\subsection{Model and Notation}
\label{sec:sub:models}

The Hamiltonian of the anharmonic oscillator under study is
in first quantization
\begin{equation}
  \label{eq:ham_quartic}
  H=\frac{1}{2}\left(P^2+\omega^2 X^2\right) + \lambda X^4\ ,
\end{equation}
where $\lambda$ is a positive constant. We will consider two cases. For 
$\omega=0$, we consider the pure quartic oscillator (PQO) for which  
 a rescaling $X\to X/\alpha$, $P\to \alpha P$ and 
$H\to H/\alpha^2$ leads to a unique Hamiltonian. 
Choosing $\alpha^6=6\lambda$ it reads
\begin{equation}
  \label{eq:ham_pqo}
  H=\frac{1}{2}P^2 + \frac{1}{6} X^4\ .
\end{equation}
For $\omega\neq 0$, we use another rescaling to bring the Hamiltonian in 
the form
\begin{equation}
  \label{eq:ham_qo}
  H=\frac{1}{2}\left(P^2+X^2\right) + \lambda X^4\ .
\end{equation}
The Hamiltonian (\ref{eq:ham_qo}) will be called QO in the sequel.

Both Hamiltonians have been studied numerically, see for instance Ref.\ 
\onlinecite{hioe78}.
Nowadays, they can be diagonalized numerically on a workstation in
 seconds, so that we  can easily check the quality of our approximations.

For both Hamiltonians, we introduce annihilation and creation operators
\begin{subequations}
  \label{eq:a_et_adag_sigma}
  \begin{eqnarray}
    a(\sigma) &=& 
    \frac{1}{\sqrt{2}}\left(\frac{X}{\sigma}+\rmi\sigma P\right)\\
    \adag(\sigma) &=& \frac{1}{\sqrt{2}}\left(\frac{X}{\sigma}-\rmi\sigma 
    P\right)\ ,
  \end{eqnarray}
\end{subequations}
where $\sigma$ is a free parameter which will be determined later. Recall that 
the QO reduces to the conventional harmonic oscillator for $\lambda=0$. It is
diagonalized by using $\sigma=1$ in Eq.\ (\ref{eq:a_et_adag_sigma}). The 
Hamiltonians will be used in their normal-ordered form. Normal-ordering is done
with respect to the bosonic vacuum which depends on the value of $\sigma$. 
The $\adag$ operators are put to the left of the $ a$ operators. 
Thus the general form of the Hamiltonians reads  
\begin{eqnarray}
  \label{eq:ham_aadag}
  H&=&g_{0,0}+g_{0,1}\adag \ana+g_{0,2}\adag^2 \ana^2\\
  &&+\adag^2\left(g_{2,0}+g_{2,1}\adag \ana\right)+\left(g_{2,0}+g_{2,1}
  \adag \ana\right) \ana^2\nonumber\\
  &&+\adag^4 g_{4,0}+g_{4,0} \ana^4\ .\nonumber
\end{eqnarray}
For brevity of the notation, we do not denote the explicit dependence of
the couplings and operators on  $\sigma$ in the above equation.
 The coupling constants $g_{i,j}$ have two indices which indicate the 
powers of $\adag$ and $ a$ in the corresponding operator.
For instance, $g_{i,j}$ is associated to the operator
$\adag^i \cdot \adag^j \ana^j$ and to its hermitian conjugate. 

For illustration we consider the harmonic oscillator (\ref{eq:ham_bos_quad}).
The ground state energy is given by $E=g_{0,0}$, 
the single-particle energy by $\omega=g_{0,1}$, and the
Bogoliubov term by $d/2=g_{2,0}$. For the PQO,
the bare values of the coupling constants are
\begin{subequations}
  \label{eq:bare_coup_pqo}
  \begin{eqnarray}
    g_{0,0}^{(\rmb)}(\sigma) &=&
    \frac{1}{4}\left( \frac{1}{\sigma^2}+\frac{1}{2}
    \sigma^4\right)\\
    g_{0,1}^{(\rmb)}(\sigma) &=& \frac{1}{2}\left( 
    \frac{1}{\sigma^2}+\sigma^4\right)
    \\
    g_{0,2}^{(\rmb)}(\sigma) &=& \frac{1}{4}\sigma^4\\
    g_{2,0}^{(\rmb)}(\sigma) &=& 
    \frac{1}{4}\left( -\frac{1}{\sigma^2}+\sigma^4\right)
    \\
    g_{2,1}^{(\rmb)}(\sigma) &=& \frac{1}{6}\sigma^4\\
    g_{4,0}^{(\rmb)}(\sigma) &=& \frac{1}{24}\sigma^4\ .
  \end{eqnarray}
\end{subequations}
For the QO, they are found to be
\begin{subequations}
  \begin{eqnarray}
    \label{eq:bare_coup_qo}
    g_{0,0}^{(\rmb)}(\sigma) &=& \frac{1}{4}\left(\frac{1}{\sigma^2}+\sigma^2
    +3\lambda\sigma^4\right)\\
    g_{0,1}^{(\rmb)}(\sigma) &=& \frac{1}{2}\left(\frac{1}{\sigma^2}+\sigma^2
    +6\lambda\sigma^4\right)\\
    g_{0,2}^{(\rmb)}(\sigma) &=& \frac{3}{2}\lambda\sigma^4\\
    g_{2,0}^{(\rmb)}(\sigma) &=& \frac{1}{4}\left( -\frac{1}{\sigma^2}+\sigma^2
    +6\lambda\sigma^4\right)\\
    g_{2,1}^{(\rmb)}(\sigma) &=& \lambda\sigma^4\\
    g_{4,0}^{(\rmb)}(\sigma) &=& \frac{1}{4}\lambda\sigma^4\ .
  \end{eqnarray}
\end{subequations}

\subsection{Perturbative Results}
\label{sec:sub:prel_res}

We start by analyzing the QO by perturbative schemes. The results
shall serve as reference for the following more sophisticated approaches.

\begin{figure}[htbp]
  \centering
  \includegraphics[width=8cm]{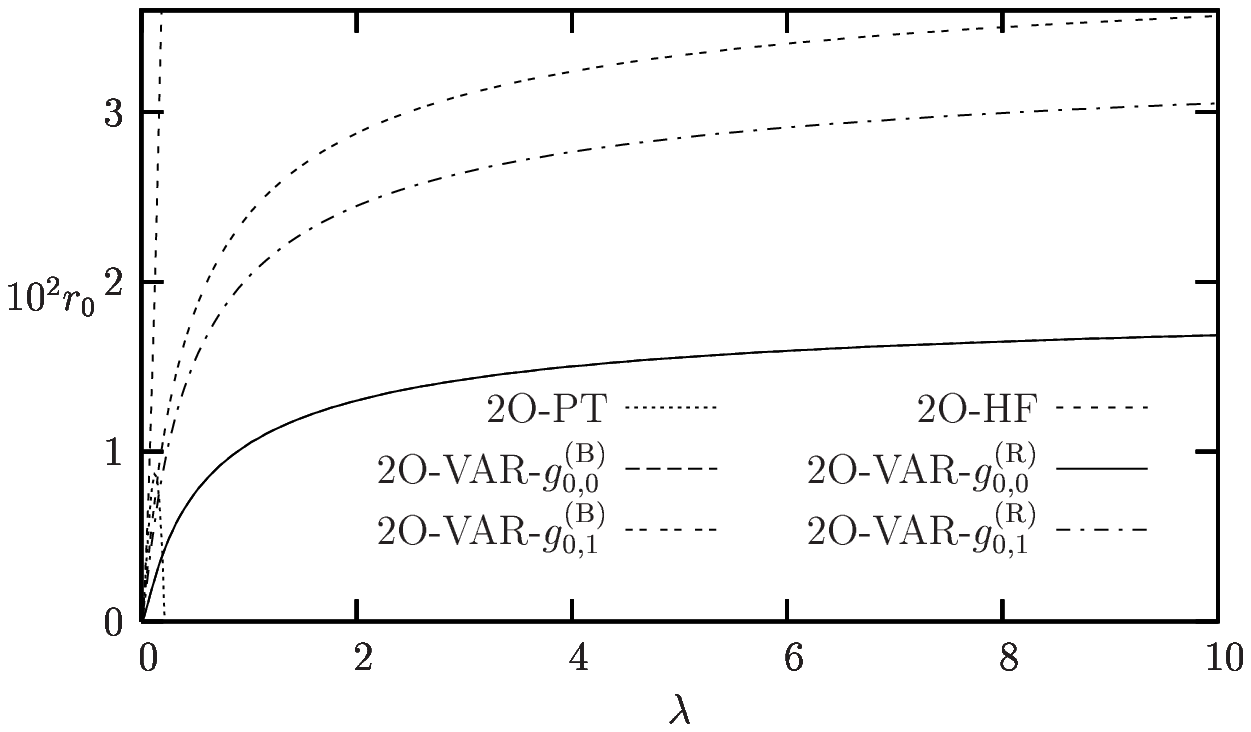}
  \includegraphics[width=8cm]{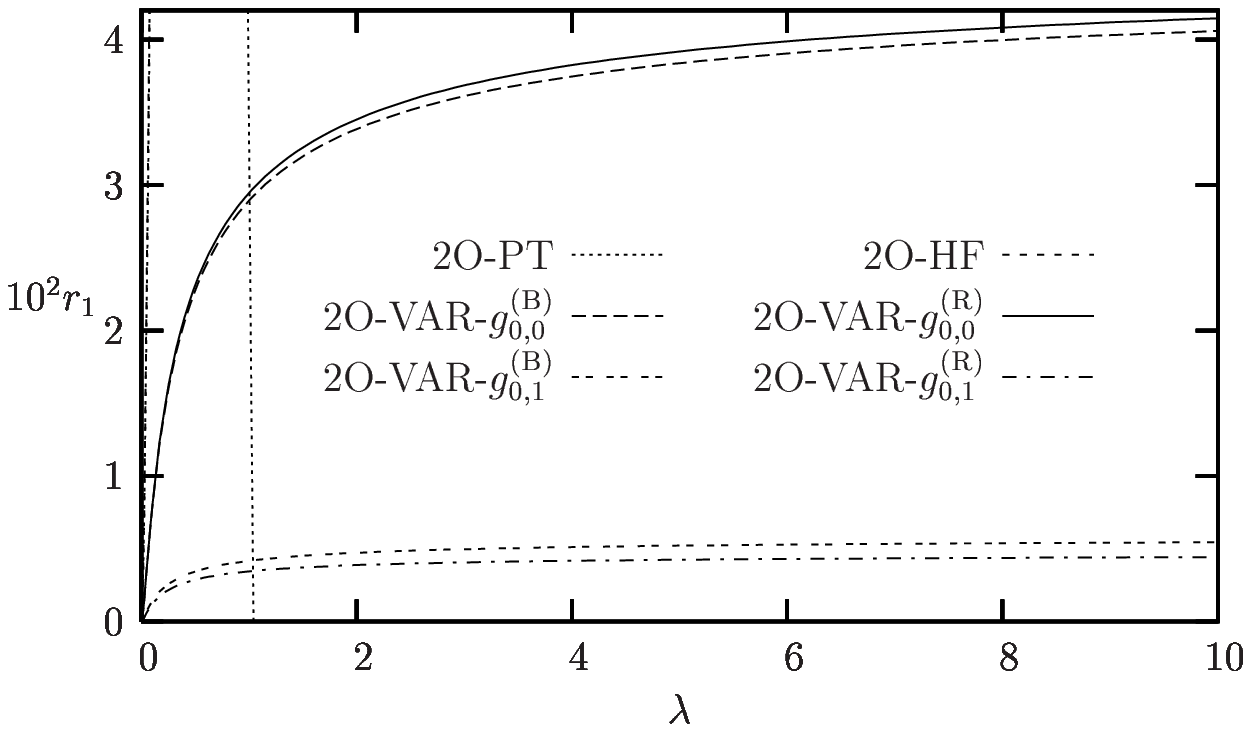}
  \includegraphics[width=8cm]{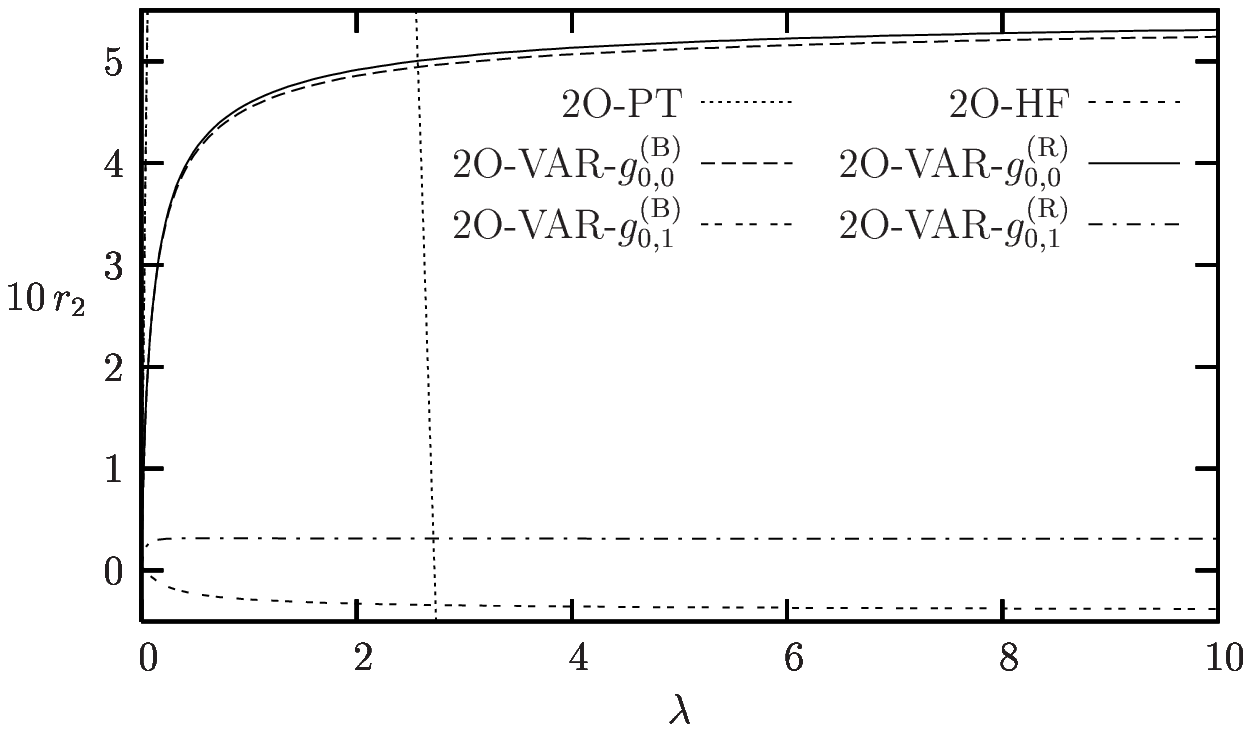}
  \caption{Relative errors $r_0$, $r_1$ and $r_2$ of the first three diagonal 
    coefficients $g_{0,0}^{(\rmr)}$, $g_{0,1}^{(\rmr)}$, $g_{0,2}^{(\rmr)}$
    (see text) for five different computation schemes. For the second 
    order perturbation theory (2O-PT) $r_1$ reaches a maximum value 
    $\simeq 23\%$ for $\lambda\sim 0.5$ (outside of the graph) and then 
    decreases again; $r_2$ reaches a maximum values $\simeq 300\%$ for 
    $\lambda\sim 1.1$. For the second order calculation around the Hartree-Fock
    Hamiltonian (2O-HF), all $r_i$ grow very fast. The results of the second 
    order calculation around the optimum bare value of the ground state energy 
    (2O-VAR-$g_{0,0}^{(\rmb)}$) and the one around the optimum value of the 
    renormalized $g_{0,0}^{(\rmr)}$ (2O-VAR-$g_{0,0}^{(\rmr)}$) are always 
    very close to each other. For $r_0$ they cannot be distinguished.}
  \label{fig:SOPT}
\end{figure}

The first approach is to expand perturbatively in 
$\lambda$ around the diagonal Hamiltonian  for $\sigma=1$. 
It is known that the resulting series is not convergent but asymptotic
\cite{hioe78}. Yet
the results may approximate the exact ones if they are restricted to low orders
in $\lambda$. In this sense, we computed the first three 
renormalized eigen energies $E_0^{(\rmr)}$, $E_1^{(\rmr)}$ and $E_2^{(\rmr)}$ 
to second order in $\lambda$. Their bare values are 
$E_0^{(\rmb)}=g_{0,0}^{(\rmb)}$, 
$E_1^{(\rmb)}=g_{0,0}^{(\rmb)}+g_{0,1}^{(\rmb)}$ and 
$E_2^{(\rmb)}=g_{0,0}^{(\rmb)}+2g_{0,1}^{(\rmb)}+2g_{0,2}^{(\rmb)}$.
{}From the approximate first three eigen energies we deduce the  renormalized 
diagonal coefficients of the Hamiltonian 
$g_{0,0}^{(\rmr)}$, $g_{0,1}^{(\rmr)}$, $g_{0,2}^{(\rmr)}$. Fig.~\ref{fig:SOPT}
displays the relative errors of these coefficients compared to the exact 
values, \ie, $r_i=\left[g_{0,i}^{(\rmr)}-g_{0,i}^{(\mathrm{ex})}\right]/
g_{0,i}^{(\mathrm{ex})}$. The curves are labelled 2O-PT for second order 
perturbation  theory. As expected the errors grow very quickly on increasing 
$\lambda$.

Instead of passing tediously to higher orders of $\lambda$ we choose another
route. We consider all terms which preserve the number of quanta as diagonal.
This means all terms with equal number of creation and annihilation operators
are kept in the diagonal Hamiltonian. Then we perform second order perturbation
theory around this diagonal part in all the off-diagonal contributions, \ie,
the terms which change the number of bosonic quanta. For concreteness, we 
give the resulting expression for the ground-state energy 
\begin{eqnarray}
  \label{eq:g00_2OPT}
  g_{0,0}^{(\rmr)}(\sigma)&=&g_{0,0}^{(\rmb)}(\sigma)-
\frac{{g_{2,0}^{(\rmb)}}^2(\sigma)}{2(g_{0,1}^{(\rmb)}(\sigma)
+g_{0,2}^{(\rmb)}(\sigma))}\\
  &&\hspace{1.5cm}-\frac{{g_{4,0}^{(\rmb)}}^2(\sigma)}
{4(g_{0,1}^{(\rmb)}(\sigma)+3g_{0,2}^{(\rmb)}(\sigma))}\ .\nonumber
\end{eqnarray}
Now, we exploit the freedom to choose $\sigma$. Five choices are studied:
\begin{itemize}
\item $\sigma_\mathrm{HF}=1$.  This choice diagonalizes the harmonic part of 
  the Hamiltonian (\ref{eq:ham_qo}). Hence we denote this scheme 2O-HF since it
  represents  second order perturbation around the  Hartree-Fock Hamiltonian.
\item $\sigma$ is chosen to fulfill 
  $\partial_\sigma g_{0,0}^{(\rmb)}(\sigma)=0$, \ie, a stationary point of the 
  bare approximate ground state energy is chosen. It can be checked that this
  is the same as to have $g_{2,0}(\sigma)=0$ vanish. So the 
  bilinear part of the whole bare Hamiltonian is diagonal. This scheme is 
  denoted by 2O-VAR-$g_{0,0}^{(\rmb)}$ since it represents second order 
  perturbation  around the stationary point of $g_{0,0}^{(\rmb)}$. 
  Conventionally, this choice is called second order perturbation around the 
  self-consistent Hartree-Fock Hamiltonian.
\item In analogy to the previous  choice  we require that the bare 
  {\em single-particle energy} is stationary $\pal g_{0,1}^{(\rmb)}=0$. This 
  choice is called   2O-VAR-$g_{0,1}^{(\rmb)}$.
\item Instead of looking for stationary points of the bare parameters, we can
  look for the stationary points of renormalized parameters. We denote by 
  2O-VAR-$g_{0,0}^{(\rmr)}$ the second order perturbation around the 
  stationary point of the renormalized ground state energy 
  $\partial_\sigma g_{0,0}^{(\rmr)}(\sigma)=0$.
\item We call 2O-VAR-$g_{0,1}^{(\rmr)}$ the analogous variation of the 
  renormalized single-particle energy.
\end{itemize}

On the mean-field level it is well-established to look for stationarity
in order to define the optimum choice of parameters. To do the
same for higher order results, here second order, is justified by the
argument that in an exact calculation the result does {\em not}
depend on $\sigma$. The value of $\sigma$ only defines a starting point
of the calculation. If the calculation is exact it does not depend on
the starting point. Hence it is stationary with respect to the variation
of the parameters defining the starting point. An approximate treatment, 
however, will not show this independence on the starting point. The best that
can be achieved is stationarity \cite{steve81} which therefore defines 
the optimum starting point.

The results obtained for these five choices are represented in  
Fig.~\ref{fig:SOPT}. Clearly, the 2O-HF result leads to no improvement compared
to the 2O-PT result. The four other choices using a variational criterion lead 
to much better results. Apart from the ground state energy where the 
2O-VAR-$g_{0,0}^{(\rmr)}$ scheme is the best, the overall best choice is the 
2O-VAR-$g_{0,1}^{(\rmr)}$ scheme, especially for 
$r_2$. Note that there is no big difference between 
2O-VAR-$g_{0,0}^{(\rmr)}$ and 2O-VAR-$g_{0,0}^{(\rmb)}$. On average, the 
2O-VAR-$g_{0,0}^{(\rmr)}$ scheme computes the $g_{0,i}^{(\rmr)}$ coefficients
 with an accuracy ranging from half a percent to a few percents. Thus we
conclude from  these calculations that the use of a variational criterion 
is always very rewarding. We will also use such a criterion to choose
the starting point of the CUT approaches.

Finally, we address the PQO without any harmonic part. It represents the
$\lambda \to \infty$ limit of the QO considered before. Of course,
neither a direct perturbative approach nor a Hartree-Fock approach is defined.
But all the other approaches can be used for the PQO without conceptual
difficulty. The results are given in Table \ref{tab:SOPT}. 
\begin{table}[htbp]
  \begin{tabular}{|c|c|c|c|}
    \hline
    & $10^2 r_0$ & $10^2 r_1$ & $10^2 r_2$ \\
    \hline\hline
    2O-VAR-$g_{0,0}^{(\rmb)}$ & 1.94 & 4.47 & 54.5 \\
    \hline
    2O-VAR-$g_{0,0}^{(\rmr)}$ & 1.94 & 4.57 & 55.3 \\
    \hline
    2O-VAR-$g_{0,1}^{(\rmb)}$ & 4.01 & 0.59 & -4.07 \\
    \hline
    2O-VAR-$g_{0,1}^{(\rmr)}$ & 3.45 & 0.47 & 3.08 \\
    \hline
  \end{tabular}
  \caption{Comparison of the four variational second order calculations for the
    purely quartic oscillator.  
    \label{tab:SOPT}}
\end{table}
Again, it turns out  that the stationary point of the single-particle energy 
leads to the best results. The error in the interaction value analyzed by 
$r_2$ is reduced by  a factor of 10 compared to the stationary point of the 
ground state energy. The deviation in the ground state energy
is smallest at its stationary points, but only
doubled when the single-particle energy is made stationary.


\subsection{Non-Perturbative Results}
\label{sec:sub:nonpert_res}

\subsubsection{Spectrum}
\label{sec:sub:spectrum}

Let us consider the Hamiltonian given in Eq.~(\ref{eq:ham_aadag}). Performing
an infinitesimal unitary transformation generates some new terms in the 
Hamiltonian which were not present before. Focusing, however, on the states
at low energies the new terms with at least six bosonic operators
will be of minor importance. They represent processes which involve at
least three particles. Hence we discard the new terms for the moment once
they have been brought into normal-ordered form. Recall that on 
normal-ordering higher order terms, terms of lower order occur. These
are kept since they matter for the low-lying states.

If the flowing Hamiltonian is truncated in this way it stays in the form of 
Eq.~(\ref{eq:ham_aadag}). It is natural to view all terms conserving the number
of bosons, namely $g_{0,0}+g_{0,1}\adag \ana+g_{0,2}\adag^2 \ana^2$, as 
the diagonal part $H_\rmd$. The non-conserving term make up the 
non-diagonal part of the Hamiltonian. With this choice, Wegner's generator 
$\ew=[H_\rmd,H_\mathrm{nd}]$ is computed. It is found to comprise terms of the 
form $\adag^4 \ana^2-\adag^2 \ana^4$ and $\adag^5 a-\adag \ana^5$ which are not
present in the initial Hamiltonian. Hence we omit them also in the 
generator\footnote{We performed also computations where these terms are 
kept in the generator.  The ensuing results are worse than those obtained 
without these terms.} which reads
\begin{eqnarray}
  \label{eq:gen}
  \eta &=&\eta_{2,0}(\adag^2- \ana^2) + \eta_{2,1}(\adag^3 a-\adag \ana^3) \\
  &&\hspace{3.5cm}+ \eta_{4,0}(\adag^4- \ana^4),\nonumber
\end{eqnarray}
where for Wegner's generator
\begin{subequations}
  \label{eq:gen_W}
  \begin{eqnarray}    
    \eta_{2,0} &=& 2(g_{0,1}+g_{0,2})g_{2,0}\\
    \eta_{2,1} &=& 2(g_{0,1}+3g_{0,2})g_{2,1}+4g_{0,2}g_{2,0}\\
    \eta_{4,0} &=&4(g_{0,1}+3g_{0,2})g_{4,0}\ .
  \end{eqnarray}
\end{subequations}
For the MKU-generator the coefficients can be read off from the 
Hamiltonian
\begin{subequations}
  \label{eq:gen_MKU}
  \begin{eqnarray}    
    \eta_{2,0} &=& g_{2,0}\\
    \eta_{2,1} &=& g_{2,1}\\
    \eta_{4,0} &=& g_{4,0}\ .
  \end{eqnarray}
\end{subequations}
It is one of the advantages of the MKU-generator that it allways has a simpler 
form than Wegner's generator, as exemplified here for the quartic oscillator. 
This was not obvious for the quadratic oscillator or the two-level system 
which are too simple to exhibit this general feature.
The flow equations read
\begin{subequations}
  \begin{eqnarray}
    \label{eq:flow_eq}
    \pal g_{0,0}&=&-4\eta_{2,0}g_{2,0}-48\eta_{4,0}g_{4,0}\\
    \pal g_{0,1}&=&
    -4\eta_{2,0}(2g_{2,0}+3g_{2,1})-12\eta_{2,1}(g_{2,0}+g_{2,1})
    \nonumber\\
    &&\hspace{2cm}-192\eta_{4,0}g_{4,0}\\
    \pal g_{0,2}&=&-12\eta_{2,0}g_{2,1}-12\eta_{2,1}(g_{2,0}+3g_{2,1})\\
    &&\hspace{2cm}-144\eta_{4,0}g_{4,0}\nonumber\\
    \pal g_{2,0}&=&-2\eta_{2,0}(g_{0,1}+g_{0,2}+6g_{4,0})-24\eta_{2,1}g_{4,0}
    \nonumber\\
    &&\hspace{2cm}-12\eta_{4,0}(g_{2,0}+2g_{2,1})\\
    \pal g_{2,1}&=&-4\eta_{2,0}(g_{0,2}+2g_{4,0})
    -4\eta_{4,0}(2g_{2,0}+9g_{2,1}) \nonumber\\
    &&-2\eta_{2,1}(g_{0,1}+3g_{0,2}+18g_{4,0})\\
    \pal g_{4,0}&=&-2\eta_{2,0}g_{2,1}+2\eta_{2,1}g_{2,0}\\
    &&\hspace{2cm}-4\eta_{4,0}(g_{0,1}+3g_{0,2})\ .\nonumber
  \end{eqnarray}
\end{subequations}
The derivation of these expressions is presented in Appendix \ref{app:flow_eq}.

We start with the results for the PQO. All the results for the eight possible 
computational schemes are collected in Table \ref{tab:CUT}. The eight 
schemes are made up from two possible generators (W or MKU) and from the
four choices which quantity is made stationary, see previous section.
\begin{table}[htbp]
  \begin{tabular}{|c|c|c|c|}
    \hline
    & $10^2r_0$ & $10^2r_1$ & $10^2r_2$ \\
    \hline\hline
    W-VAR-$g_{0,0}^{(\rmb)}$ & 0.127 & 0.000876 & -15.8 \\
    \hline    
    W-VAR-$g_{0,0}^{(\rmr)}$ & -0.0815 & -0.195 & -1.86 \\
    \hline
    W-VAR-$g_{0,1}^{(\rmb)}$ & -0.0612 & -0.279 & -4.83 \\
    \hline
    W-VAR-$g_{0,1}^{(\rmr)}$ & -0.0426 & -0.287 & -6.22 \\
    \hline
    MKU-VAR-$g_{0,0}^{(\rmb)}$ & 0.0492 & -0.136 & -16.5 \\
    \hline
    MKU-VAR-$g_{0,0}^{(\rmr)}$ & -0.0608 & -0.232 & -0.951 \\
    \hline
    MKU-VAR-$g_{0,1}^{(\rmb)}$ & -0.0555 & -0.289 & -3.51 \\
    \hline    
    MKU-VAR-$g_{0,1}^{(\rmr)}$ & -0.0352 & -0.315 & -6.97 \\
    \hline
  \end{tabular}
  \caption{Comparison of the four variational CUT calculations for the PQO 
     with Wegner's or MKU generator.  
    \label{tab:CUT}}
\end{table}
We have to mention that $g_{0,0}^{(\rmr)}$ and $g_{0,1}^{(\rmr)}$ display both
a local minimum and a local maximum, \ie, there are two
choices for the stationary point. Here only  the local minima are considered.

The comparison of Tables \ref{tab:SOPT} and \ref{tab:CUT} reveals that the CUT 
results are always much more accurate than the second order 
calculations performed in Sec.~\ref{sec:sub:prel_res}. This is of course
expected and does not depend on  the chosen variational scheme.
For instance, the ground state energy is mostly given to an accuracy of less 
than 0.1\%, to be compared to the 2\% of the second order calculations. 
Yet the calculation is straightforward. Only six differential
equations need to be solved which can be done by any computer algrebra
program. This evidences that the CUTs lead to a significant improvement.

To vary the renormalized quantities instead of the bare ones modifies the
CUT results  more than it modified
the second order results in Table \ref{tab:SOPT}.
In the CUT calculations, the stationary points of the ground state energy 
provide the best results. This is in contrast to the finding in the 
perturbative results where the stationary
points of the single-particle energy were optimum.
We stress that the variation of the renormalized quantities is superior
to the variation of the bares one if the focus is set not only on the ground
state energy but in particular to the single-particle and two-particle
energies.

In comparison, the results for the two choices of the generator are
very similar. The MKU-generator works a little better, for instance
for the variation of the renormalized ground state energy. We attribute
this to the fact that it avoids the appearance of terms which create
 or annihilate more than four bosons. Hence there are some terms which
are not created so that they do not need to be neglected in the truncation 
scheme. But this effect is obviously fairly small for the system under study.

From these observations we will restrict our computations for the
QO to the schemes which are based on stationarity, \ie, we do not show the
direct Hartree-Fock calculation. We  performed such calculations and the 
results were unsatisfactory. The flow equations did not even converge to a 
diagonal Hamiltonian fixed point for $\lambda\gtrsim 4$. This illustrates
again that the starting point should not be too 
far away from the exact result.
\footnote{A particularly
drastic illustration of this statement is the attempt to derive the
spectrum of the harmonic oscillator starting from free particles,
even though no truncations are necessary.
Consider $H=(P^2+\delta\omega^2 X^2)/2$ with $\delta\omega^2$ being small.
One may  try to choose the $P^2$ term as the diagonal part of the 
Hamiltonian and the  $X^2$ term as the non-diagonal part. Writing 
$H(l)=[\alpha(l) P^2+\beta(l)X^2]/2$, it is found that 
$\ew=-2\rmi\alpha\beta(XP+PX)$ so that the flow equations 
$\pal \alpha=2\alpha^2\beta$ and $\pal \alpha=-2\alpha^2\beta$ result. 
Thus $\alpha\beta$ is a constant, and as 
$\beta\to 0$, $\alpha\to\infty$. Of course, this was to be expected
since the spectrum of an harmonic oscillator is discrete so that it
cannot be mapped one-to-one to the continuous spectrum of $P^2$.}
\begin{figure}[htbp]
  \centering
  \includegraphics[width=8cm]{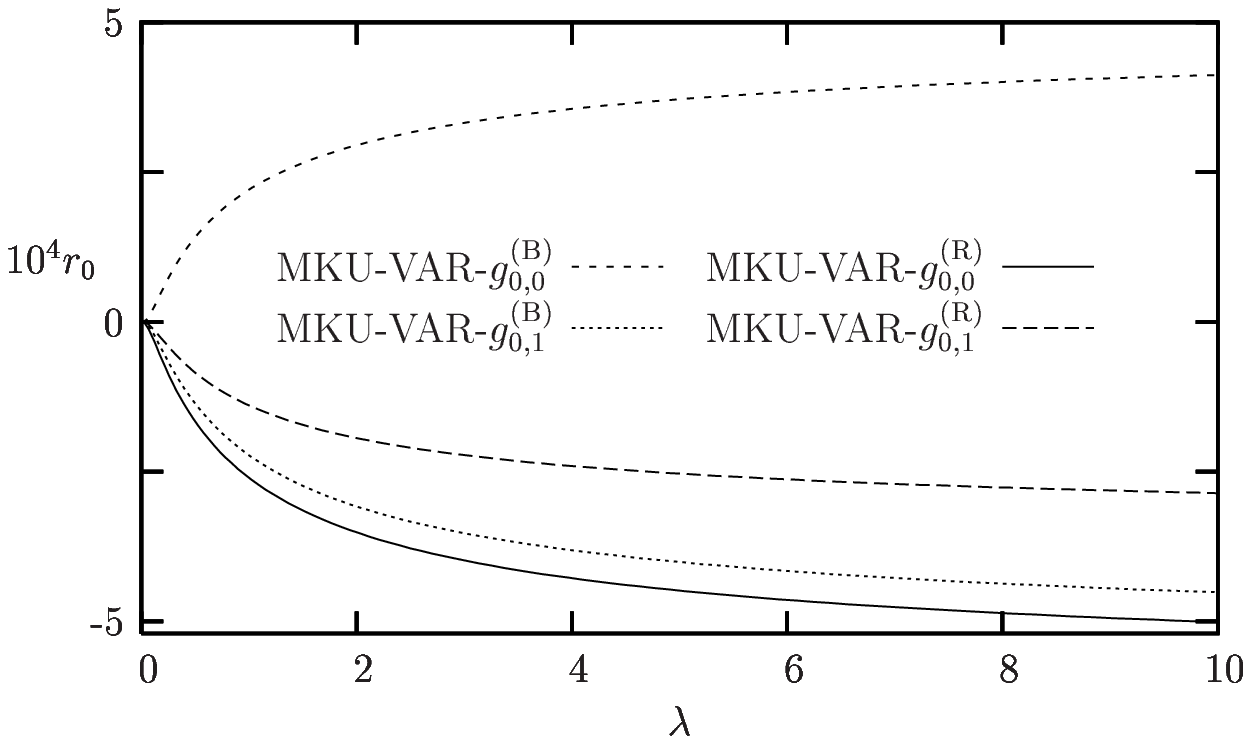}
  \includegraphics[width=8cm]{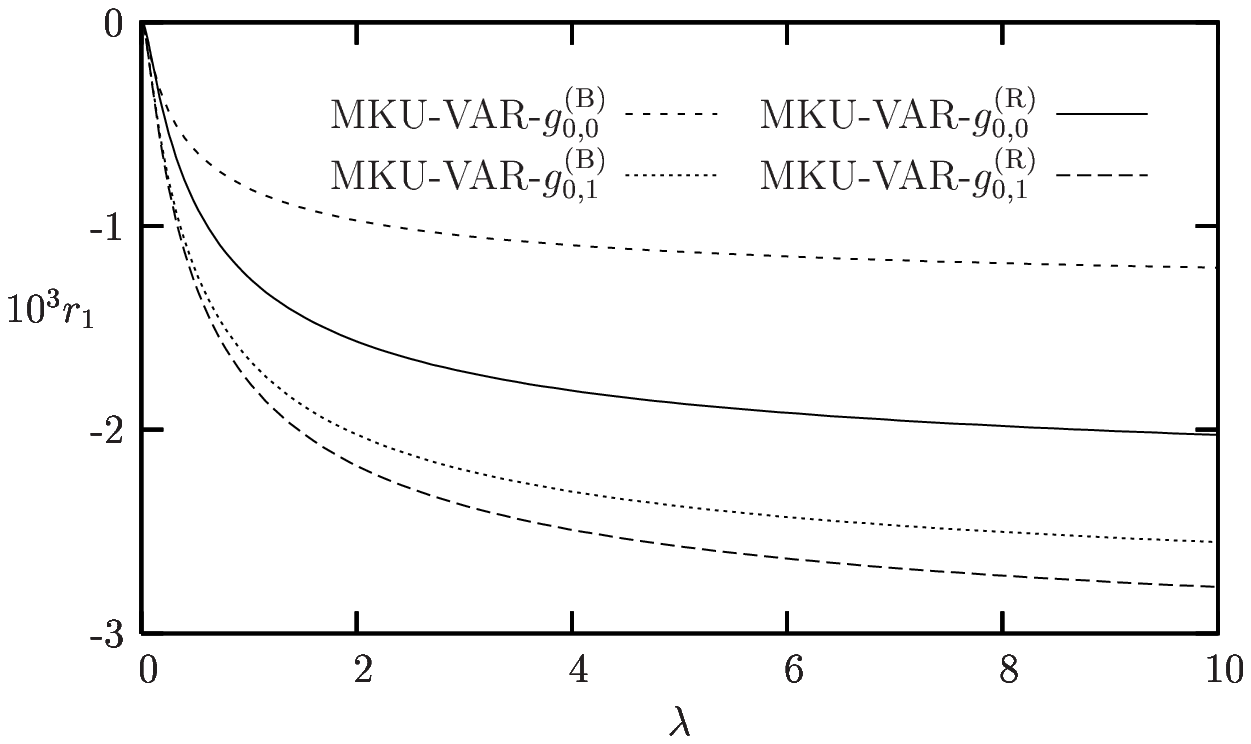}
  \includegraphics[width=8cm]{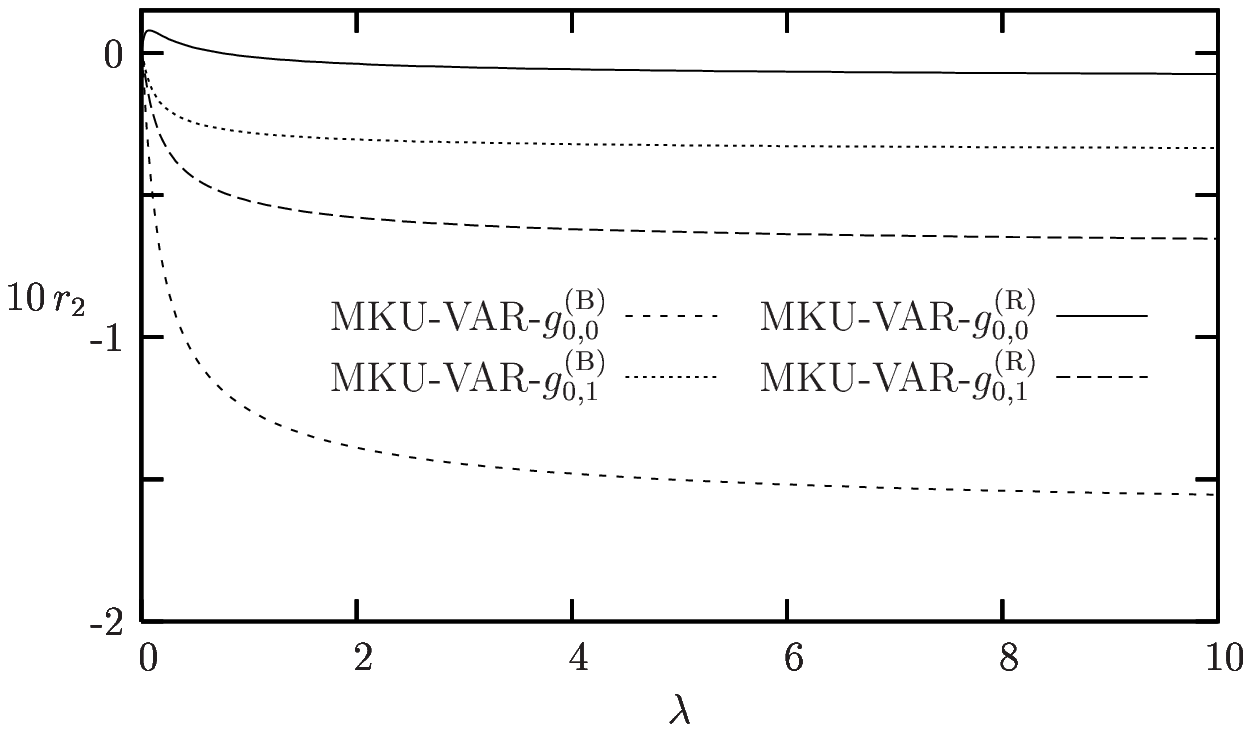}
  \caption{Relative errors $r_0$, $r_1$ and $r_2$ of the first three diagonal 
    coefficients $g_{0,0}^{(\rmr)}$, $g_{0,1}^{(\rmr)}$, $g_{0,2}^{(\rmr)}$
    (see text) for four variation schemes based on the MKU-generator.}
  \label{fig:MKU}
\end{figure}

Furthermore, we show results only for the MKU-generator. Its results are 
representative for similar calculations based on Wegner's generator. 
They are shown in Fig.\ \ref{fig:MKU}. Moreover, the MKU-generator
is easier to implement 
(compare Eqs.~(\ref{eq:gen_W}) and (\ref{eq:gen_MKU}) above).
This fact will be of importance for calculations
comprising more terms (see below).

The accuracy of the results is very impressive in view of the simplicity
of the numerical effort. As for the PQO the variation
of the renormalized, \ie, final ground state energy yields the best
results if the focus includes the two-particle energies whose accuracy
is given by $r_2$.

Finally, a remark about another possible choice of the generator is in order
since its generalisations  have been used in more 
complex physical situations, e.g.\ in Refs.\ \onlinecite{kehre96a, kehre96b}.
Using the W-VAR-$g_{0,0}^{(\rmb)}$ scheme, the quadratic 
off-diagonal coefficient $g_{2,0}$ vanishes at $l=0$. But it will become
finite in the course of the flow. It is possible to modify 
Wegner's generator to keep this coefficient zero during the flow. To this
 end, the $\eta_{2,0}$ coefficient of the generator is changed.
It is chosen to be
$\eta_{2,0}=-72g_{2,1}g_{4,0}(g_{0,1}+3g_{0,2})/
(g_{0,1}+g_{0,2}+6g_{4,0})$ instead of $2(g_{0,1}+g_{0,2})g_{2,0}$. 
This choice has two major advantages. First, the coefficient $g_{2,0}$
does not need to be traced. Second, we find that it converges faster to its
diagonal Hamiltonian. For the PQO, the accuracies are found to be
$10^2r_0=0.00569$, $10^2r_1=1.53$ and $10^2r_2=-13.6$. 
Thus we obtain the best value of the ground state energy with this scheme 
but also the worst value of the single-particle energy. 
The value of the interaction is as inaccurate as it is for the
W-VAR-$g_{0,0}^{(\rmb)}$ scheme. Furthermore, we have  lost the possibility
to vary the initial value of the parameter $\sigma$ since it is fixed to 
unity by the condition $g_{2,0}=0$. An adapted choice of $\sigma$
will be important in dealing with more terms so that we consider
it a caveat that $\sigma$ is already fixed.
Summarizing, this variant yields very reasonable results, 
but it is not the optimum choice.


\subsubsection{Spectral Weights}
\label{sec:sub:spectral_weights}

Apart from the spectrum of the oscillator, one is also interested in the 
computation of the Green function and of the spectral weights.
The retarded Green function is defined as
\begin{equation}
  \label{eq:green_realtime}
  G(t)=-i\langle 0|\left[ X(t), X(0)\right] |0\rangle \Theta(t),
\end{equation}
where $X(t)=\exp(\rmi Ht)X\exp(-\rmi Ht)$, $|0\rangle$ is the ground state. 
The Fourier transform of the Green function reads
\begin{eqnarray}
  \label{eq:green_fourier}
  G(\omega)&=&\int_{-\infty}^{+\infty}\rmd t G(t)\exp(\rmi\omega t)\\
  &=&\sum_n \left| \langle n|X|0\rangle \right|^2
  \left( \frac{1}{\omega-(E_n-E_0)+\rmi 0+}\right.\\
  &&\hspace{2.5cm}\left.-\frac{1}{\omega+(E_n-E_0)+\rmi0+}\right)\ .\nonumber
\end{eqnarray}
We have denoted  the eigenstates of the Hamiltonian by $|n\rangle$ with eigen 
energies $E_n$, \ie $H|n\rangle=E_n|n\rangle$.
We fill focus on the computation of the spectral weights
$J_n:=\left| \langle n|X|0\rangle \right|^2$.
They can be found trivially after the transformation since then
the eigen states are the states with a given number of bosons
\begin{subequations}
\begin{eqnarray}
|0\rangle &=& U(\infty)|0\rangle_0\\
|n\rangle &=& U(\infty)|n\rangle_0\\
&=& U(\infty) \frac{1}{n!}\adag^n|0\rangle_0 \ .
\end{eqnarray}
\end{subequations}
So we need the transformed observable $X(l)=U^\dagger(l)XU(l)$
 which satisfies the same flow equation as the Hamiltonian 
\begin{equation}
\pal X(l)=[\eta(l),X(l)]
\end{equation} 
with initial condition $X(0)=(\sigma/\sqrt{2})(\adag+a)$. Here again, an 
infinite hierarchy of terms appears  when $X(l)$ is commuted with $\eta(l)$.
We will restrict the calculation to the lowest non-trivial truncation scheme
\begin{eqnarray}
  \label{eq:xofl}
  &&X(l)=\alpha_{1,0}(l)\left(\adag+a\right)+\alpha_{1,1}(l)\left(\adag^2a+
  \adag \ana^2\right)\nonumber\\
  &&\hspace{2cm}+\alpha_{3,0}(l)\left(\adag^3+a^3\right).
\end{eqnarray}
The flow equations of the $\alpha$ coefficients are found to be
\begin{subequations}
  \label{eq:flow_alpha}
  \begin{eqnarray}
    \pal\alpha_{1,0}&=&-2[\eta_{2,0}(\alpha_{1,0}+\alpha_{1,1}+3\alpha_{3,0})\\
    &&\hspace{1cm}+3\eta_{2,1}\alpha_{3,0}+12\eta_{4,0}\alpha_{3,0}]\nonumber\\
    \pal\alpha_{1,1}&=&-2[\eta_{2,0}(2\alpha_{1,1}+3\alpha_{3,0})]\\
    &&-3[\eta_{2,1}(\alpha_{1,0}+2\alpha_{1,1}+6\alpha_{3,0})+12\eta_{4,0}
      \alpha_{3,0}]\nonumber\\
    \pal\alpha_{3,0}&=&-2\eta_{2,0}\alpha_{1,1}+\eta_{2,1}\alpha_{1,0}\\
    &&\hspace{1cm}-4\eta_{4,0}(\alpha_{1,0}+3\alpha_{1,1}) \ .\nonumber
  \end{eqnarray}
\end{subequations}
Note that this set of equations is linear in the $\alpha$ coefficients
since the coefficients in the generator are determined from the flow
of the Hamiltonian. So for the computation of the observables the
$g$ coefficients are given and only a linear set of differential equations,
though with non-constant coefficients, has to be solved.

\begin{figure}[htbp]
  \centering
  \includegraphics[width=8cm]{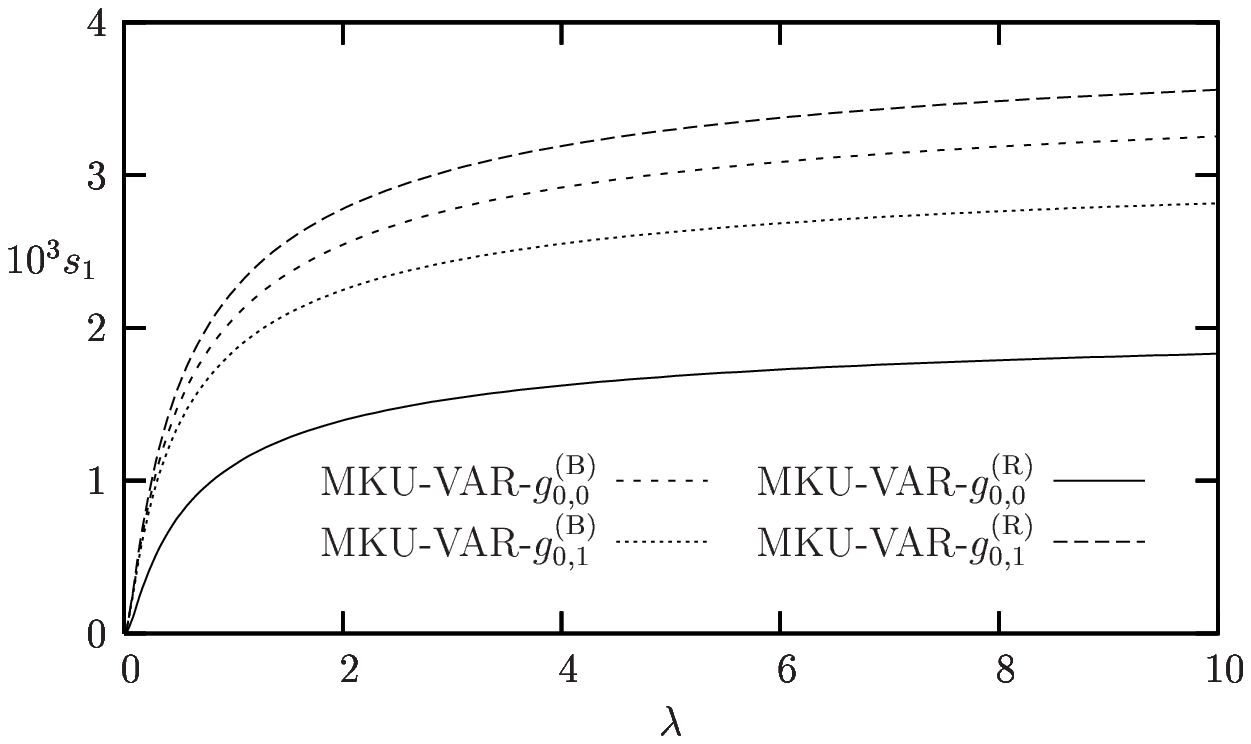}
  \includegraphics[width=8cm]{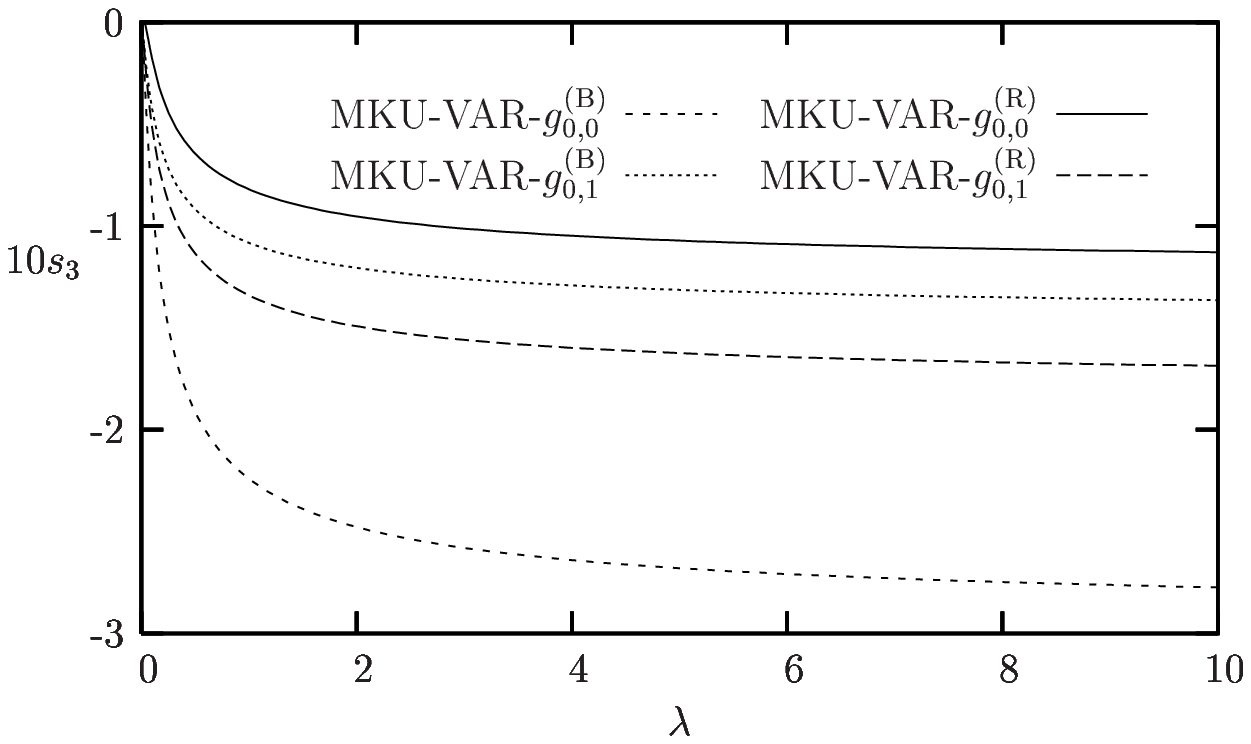}
  \caption{Relative errors $s_1$ and $s_3$ of the CUT calculation for the 
    spectral weights $J_1$ and  $J_3$ (see text) for four variation schemes 
    based on the MKU-generator.}
  \label{fig:J13}
\end{figure}

The first three $J$ coefficients can  be obtained 
from (\ref{eq:flow_alpha}); they read
$J_1=\alpha_{1,0}^2(\infty)$, $J_2=0$ due to the identical parity of the 
$|0\rangle$ and $|2\rangle$ states, and $J_3=6\alpha_{3,0}^2(\infty)$.
The parity of the states in real-space representation is expressed
in second quantization by the fact that the boson number is 
changed only by multiples of two. In Fig.~\ref{fig:J13},
the results obtained in this way, for the MKU generator, 
are compared to the numerically exact ones.
The relative errors $s_{1/3}=\left[J_{1/3}-J_{1/3}^{(\mathrm{ex})}\right]
/J_{1/3}^{(\mathrm{ex})}$ are plotted.
For $\lambda=10$, the error in $J_1$ is of the order of a few tenth of a
percent. The error on $J_3$ is about 10\%.
Note that the weight 
$J_1$ ($J_1^{(\mathrm{ex})}(\lambda=10)=1.297\cdot10^{-1}$) dominates and 
that the weight of $J_3$ ($J_3^{(\mathrm{ex})}(\lambda=10)=3.469\cdot10^{-4}$) 
is smaller by more than two orders of magnitude so that
the errors on the absolute
scale are equally small. We attribute the relatively large deviations
in $J_3$  to the low order of the truncation scheme used for the
observables and for the Hamiltonian where no three-particle energy was
included.

For the PQO, the relative errors of the computation of the spectral weights 
are given in Table \ref{tab:J13_PQO} for MKU and W generators.
\begin{table}[htbp]
  \begin{tabular}{|c|c|c|}
    \hline
    & $10^3s_1$ & $10s_3$ \\
    \hline\hline
    \hline
    W-VAR-$g_{0,0}^{(\rmb)}$ & 2.77 & -1.99 \\
    \hline
    W-VAR-$g_{0,0}^{(\rmr)}$ & 2.02 & -1.00 \\
    \hline
    W-VAR-$g_{0,1}^{(\rmb)}$ & 3.58 & -1.15 \\
    \hline    
    W-VAR-$g_{0,1}^{(\rmr)}$ & 3.96 & -1.24 \\
    \hline
    MKU-VAR-$g_{0,0}^{(\rmb)}$ & 3.71 & -2.94 \\
    \hline
    MKU-VAR-$g_{0,0}^{(\rmr)}$ & 2.12 & -1.23 \\
    \hline
    MKU-VAR-$g_{0,1}^{(\rmb)}$ & 3.17 & -1.46 \\
    \hline    
    MKU-VAR-$g_{0,1}^{(\rmr)}$ & 4.06 & -1.80 \\
    \hline    
  \end{tabular}
  \caption{Relative errors $s_1$ and $s_3$ of the CUT calculation for the 
    spectral weights $J_1$ and  $J_3$ of the PQO for four variation schemes 
    based on the W and MKU generators.}  
    \label{tab:J13_PQO}
\end{table}
The exact values for the spectral weights are $J_1^{(\mathrm{ex})}=0.521$ and 
$J_3^{(\mathrm{ex})}=0.00152$, showing that even for the PQO 
the lowest excited state has most of the spectral weight.

It is very interesting to compare the various variational schemes. The
result from the energies is confirmed. The by far best result is obtained
from the variation of the renormalized ground state energy. This fact
strongly suggests to adopt this scheme also in more complex situations.
The variation of the bare ground state energy, equivalent to the
CUT treatment around the self-consistent mean-field calculation, yields
also satisfactory results. It is easier to realize which may become
essential for extended systems. 
Finally, let us note that both generators give results which are similar, as 
was already the case for the spectrum. Wegner's generator works a little 
better for the spectral weights than the MKU generator, whereas it was the 
inverse for the spectrum.


\section{Extended Truncation Schemes}
\label{sec:ExTrSc}

In this section, we only consider the PQO of Eq.~(\ref{eq:ham_pqo}) to keep
the presentation concise. The question investigated is how the 
truncation scheme used so far can be extended to improve the results for 
the spectrum in a systematically controlled fashion. 
For simplicity, we stick to the MKU-generator.

\subsection{Structure of the Flow Equations}
\label{sec:sub:struc_flow_eq}

Using the MKU-generator, the Hamiltonian  retains its band structure during 
the flow so that no terms changing the boson number by more than four 
arise. Hence the generalisation of Eq.~(\ref{eq:ham_aadag}) reads 
\begin{eqnarray}
  \label{eq:ham_gen}
  H&=&\Big[ g_{0,i} M_i +  g_{2,i} \adag^2 M_i+g_{2,i} M_i \ana^2\\
    &&\hspace{3cm}+  g_{4,i}\adag^4 M_i+g_{4,i} M_i \ana^4\Big]\ ,\nonumber
\end{eqnarray}
where summation over the repeated index $i$ from 0 to $\infty$ is understood. 
The same convention is used in the sequel unless upper bounds for
the indices are explicitly discussed.
The operator $M_i=\adag^i \ana^i$ appearing in Eq.\ (\ref{eq:ham_gen}) 
is the normal-ordered form of the $i^\mathrm{th}$ 
power of the number operator $Q=\adag \ana$. In Wegner's scheme the 
flowing Hamiltonian would  contain terms creating or destroying 
{\em any} even number of the excitations. The MKU generator is given by
\begin{equation}
  \label{eq:eta_gen}
  \emku= g_{2,i} \adag^2 M_i-g_{2,i} M_i \ana^2 +  
  g_{4,i} \adag^4 M_i-g_{4,i} M_i \ana^4\ .
\end{equation}

The flow equations are found by commuting $\emku$ with $H$. The derivation of
these equations is included in  Appendix \ref{app:flow_eq} where also the 
relations needed for Wegner's scheme are given. 
The flow equations  have the general form
\begin{subequations}
  \begin{eqnarray}
    \label{eq:flow_eq_MKU_N}
    \pal g_{0,k} &=& -\left[ \beta\xxx{k}{0}{i}{2}{j}{2} g_{2,i} g_{2,j} + 
      \beta\xxx{k}{0}{i}{4}{j}{4} g_{4,i} g_{4,j}\right]\\
    \pal g_{2,k} &=& -\left[ \beta\xxx{k}{2}{i}{0}{j}{2} g_{0,i} g_{2,j} + 
      \beta\xxx{k}{2}{i}{2}{j}{4} g_{2,i} g_{4,j}\right]\ . \qquad\\
    \pal g_{4,k} &=& -\left[\beta\xxx{k}{4}{i}{0}{j}{4} g_{0,i} g_{4,j}\right]
  \end{eqnarray}
\end{subequations}
Let us for example consider the $g_0$'s which appear in the part of the 
Hamiltonian  conserving the number of excitations. They are 
renormalized only by the commutation of one operator creating $n=2$ (4, resp.) 
excitations and of one operator destroying $n=2$ (4, resp.) excitations.


\subsection{Truncation Schemes}
\label{sec:sub:trunc_flow_eq}

In the numerical integration of the differential equations, we can keep only a 
finite number of couplings. We denote by $m_0$, $m_2$ and $m_4$ the number of 
$g_0$, $g_2$ and $g_4$ couplings kept. For the $g_0$ couplings, this means that
we keep the couplings $g_{0,0},\ldots,g_{0,m_0-1}$. We will call the 
truncation scheme defined by the three numbers $m_0$, $m_2$ and $m_4$ the 
``$(m_0,m_2,m_4)$ scheme''. For example, the truncation scheme used in 
Sect.~\ref{sec:sub:spectrum} is the $(3,2,1)$ scheme.

One  question arising is how to choose the truncation scheme. This 
is a very difficult question which to our knowledge has no general answer. 
Starting from a Hamiltonian of the form $(m_0,m_2,m_4)$ and commuting it 
once with the corresponding MKU-generator, new terms are generated which belong
to the $(\max[2m_2,2(m_4+1)],\max[m_0+m_2-2,m_2+m_4],m_0+m_4-2)$ scheme. 
Starting from the initial $(3,2,1)$ scheme, for instance, 
the following schemes are generated
\begin{eqnarray}
  \label{eq:trun_schemes_321}
  &&(3,2,1)\to(4,3,2)\to(6,5,4)\nonumber\\
  &&\hspace{1cm}\to(10,9,8)\to(18,17,16)\to\ldots
\end{eqnarray}
These schemes are all of the type $(N,N-1,N-2)$. So we decide to consider 
the schemes of this family. Additionally, the terms of this family have the
nice property that in each class changing the number of excitations $Q$ by 
0, 2 or 4 the maximum number of creation and annihilation operators is equal
to $2(N-1)=2(m_0-1)=2+2(m_2-1)=4+2(m_4-1)$.


\subsection{Results}
\label{sec:sub:results}

\begin{figure}[htbp]
  \centering
  \includegraphics[width=8cm]{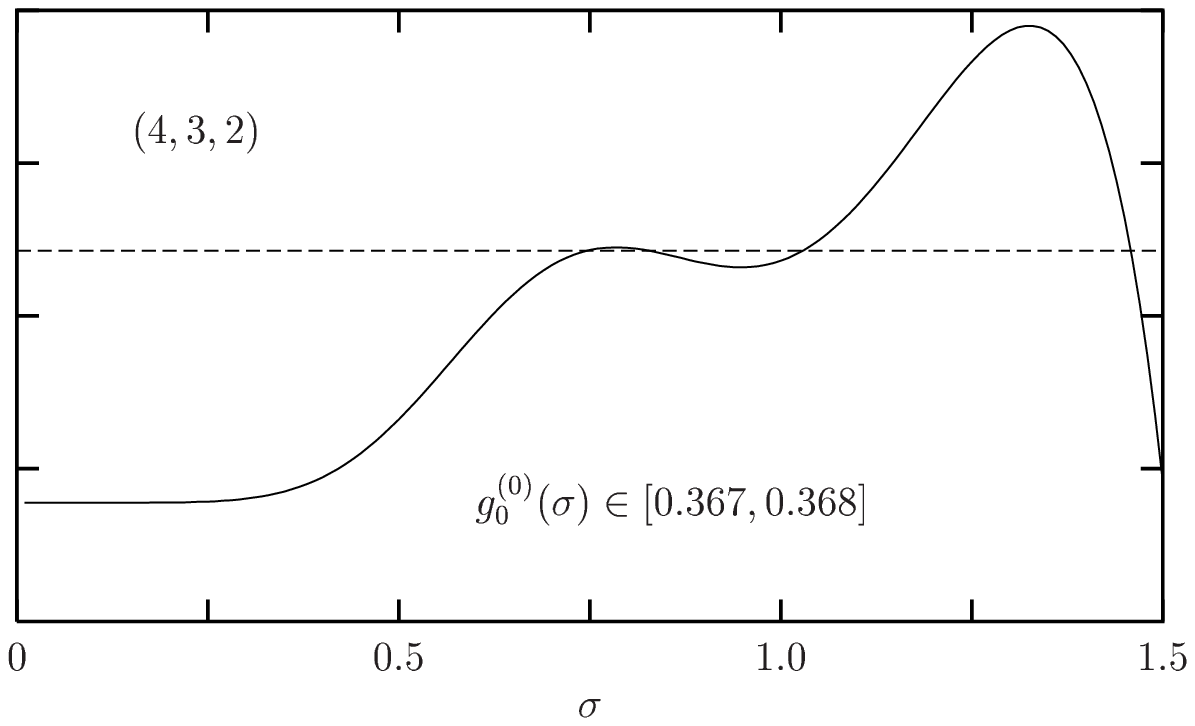}
  \includegraphics[width=8cm]{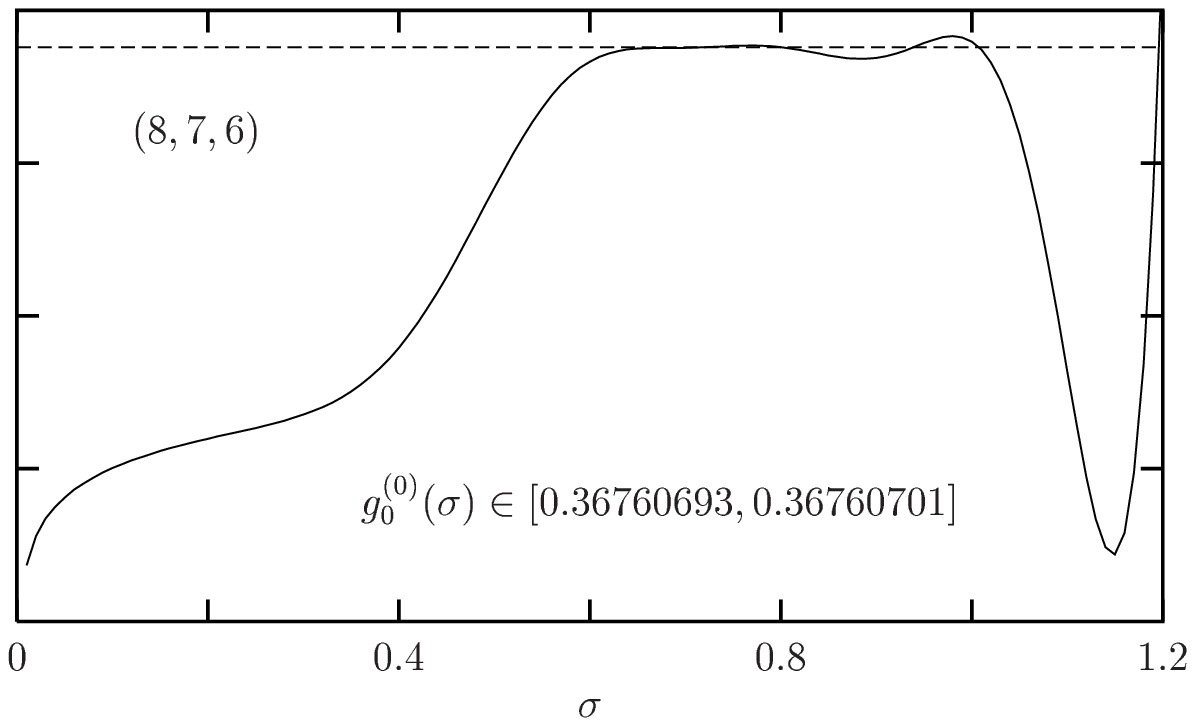}
  \caption{Ground state energy of the PQO as a function of $\sigma$ for the two
 schemes $(4,3,2)$ and $(8,7,6)$. The dashed line represents the exact ground 
state energy.}
  \label{fig:e0_n4_n8}
\end{figure}

We have checked that including more  terms in the Hamiltonian,
\ie, increasing $N$ in the $(N,N-1,N-2)$ scheme, results in a
continuously improved accuracy for the low-lying energy levels. 
In Fig.~\ref{fig:e0_n4_n8} we illustrate results for the ground-state energy
on passing from the scheme $(4,3,2)$ to the scheme $(8,7,6)$. 
For both truncation  schemes, the ground-state energy is plotted 
as a function of the parameter $\sigma$ (see Sec.~\ref{sec:QO_lowest_order}). 
Clearly, a plateau behaviour develops around a value of $\sigma$ smaller than 
1. Note in particular that the magnitude of the scale on which the 
ground state energy changes is shrinking strikingly from the 
$(4,3,2)$ to the $(8,7,6)$ scheme.
This finding agrees perfectly with our previous argument on the independence
of an exact calculation on the starting point, see Sect.\
\ref{sec:sub:prel_res}. Hence an improved calculation should approach
this independence gradually. This means that the dependence of the
quantities computed decreases as the accuracy of the calculational
schems is enhanced. This is exactly what we can observe.
The fact that values of $\sigma$ below 1 are preferred is plausible.
The $X^4$ part of the potential squeezes the eigen states more than does
the harmonic potential. 

Fig.\ \ref{fig:e0_n4_n8} shows also another feature. The results deteriorate
if  $\sigma$ becomes too big. In fact, the flows even diverge above some 
critical  value of $\sigma$. For the $(4,3,2)$ scheme this value is 
$\sigma_\rmc\simeq1.52$ and for the $(8,7,6)$ scheme
it is  $\sigma_\rmc\simeq1.21$.  These divergences originate in the error 
induced by the truncation of the flow  equations. The values of $\sigma_\rmc$ 
decrease on increasing $N$. For the values we investigated (up to 25) the 
values of $\sigma_\rmc$ remained finite. 

These results suggest that there  is no  uniform 
convergence as the limit $N\to \infty$ is taken. 
For any practical calculation, however, this is no serious limitation since
the values of $N$ cannot be chosen very large. In particular in spatially
extended systems, $N=3$ or 4 will be the maximum which can be treated.


\section{Summary and Conclusion}
\label{sec:ccl}

In this article, we presented a detailed investigation of the 
quantum quartic oscillator by perturbative and non-perturbative,
renormalizing treatments. We applied continuous unitary transformations
to determine the spectrum as well as the spectral weights of the 
quartic oscillator. This model served as the simplest interacting
quantum model. So the method could be illustrated in a simple
and transparent setting. The calculations presented may guide
future investigations of spatially extended systems like, for
instance, interacting spin waves in quantum antiferromagnets. 

We demonstrated that it is rewarding to ally the CUTs with a variational 
principle. Thereby, an optimum reference state can be chosen with respect 
to which the necessary truncations introduce only very small errors. 
It is to be expected that this is also true for other renormalizing
schemes.

In the CUT treatment presented
tracing only six couplings  achieved a remarkable accuracy. 
Without a reasonably chosen reference state, however, 
the CUT procedure does not succeed. Information on the spectral weights
complements the information needed for the Green functions. Only little
additional effort is needed to obtain the required coefficients.

We compared two generator schemes (Wegner (W), 
Eq.~(\ref{eq:wegner_generaor_general}) and Mielke-Knetter-Uhrig (MKU), 
Eq.~(\ref{eq:mku_generaor_general})).
The W generators always lead to a fixed point while the MKU generators may 
fail to converge. But the W generators are stopped by degeneracies so that 
no significant simplification is reached. 
The MKU generators are easier to write down and to implement than the W 
generators. 
In their performance they are similar for the quartic oscillator. For the 
energies the MKU is slightly better than the W generator. 
For the spectral weights it is the inverse. 
The experience with more complex, extended systems \cite{reisc04,schmi04} 
indicates that the MKU generator is very powerful if the energy eigenvalues 
are sorted in ascending order of the particle number $Q$.

Finally, we discussed how the truncation scheme for the continuous 
unitary transformation can be extended to improve the accuracy
of the approach in a systematically controlled way. Our results show
that this is indeed possible. Although no uniform convergence can be
expected the description of the low-lying states can be gradually 
improved by keeping more terms in the flow.

Ideally, our presentation will serve as seed for future investigations
of more complex systems by the approach discussed here.


\section*{ACKNOWLEDGMENTS}

We would like to thank V. Meden, A. Reischl and K.~P. Schmidt for many useful 
discussions. S.~D. also wishes to thank E. Loyer and M. Peyrard for help 
and support. Financial support of the DFG in
SP1073 is gratefully acknowledged.


\appendix

\section{Proof of convergence of Wegner's flow for infinite matrices}
\label{app:proof}

The proof of convergence given in Ref.\ \onlinecite{wegne94} is only valid for 
finite  matrices since it makes use of the invariance of the trace of 
$H^2(l)$ in  the flow. But the trace is not generally defined for
infinite systems. So the proof as given in  Ref.\ \onlinecite{wegne94} works 
for the two-level system  (\ref{eq:ham_ferm_quad}), 
but it does not work for the bosonic oscillator 
(\ref{eq:ham_bos_quad}) whose eigenvalues grow like $n$.

But the proof can be slightly modified so that it works for infinite matrices 
as well. The idea is inspired from statistical mechanics. A
regularization is needed which makes the higher lying eigen values 
contribute less. To this end, one can study $\exp(-\beta H(l))$ instead of $H$
for an arbitrary but fixed $\beta>0$. Since  $\exp(-\beta H)$ and
$H$ commute there is a certain basis in which both are diagonal.
So the aim is to diagonalize $\exp(-\beta H(l))$ which guarantees
the diagonality of $H$ up to possible degeneracies.

The generator is modified
\begin{equation}
\eta_\beta(l)=\frac{1}{\beta^2}\left[\exp(-\beta H(l))\big|_\mathrm{d},
\exp(-\beta H(l))\big|_\mathrm{nd}\right]\ .
\end{equation}
Then Wegner's proof can be used with $H$ replaced by $\exp(-\beta H)$.
The division by $\beta^2$ in $\eta$ only changes the scale of $l$. 
But it is useful since it ensures 
$\lim_{\beta\to 0} \eta_\beta=\ew$. Thus one recovers in this 
``high-temperature'' limit the generator proposed by Wegner.
Hence the flow equation 
$\pal \exp(-\beta H(l))=[\eta_\beta(l),\exp(-\beta H(l))]$ 
can be reduced to the usual one $\pal H(l)=[\ew(l),H(l)]$
because it does not matter whether the unitary transformation
is applied to the whole function or to its argument.

So the use of $\beta>0$ serves only to induce convergence.
For $\beta\to 0$ all states of the Hilbert space matter as before.


\section{Flow Equations}
\label{app:flow_eq}

\subsection{Basic Relation for Commutators of Bosonic Operators}
\label{app:sub:useful_rel}

The computation of the commutator of the generator with the Hamiltonian or the 
observables basically involves commutators $[ \ana^n,\adag^p ]$. 
It suffices to consider $n\leqslant p$ since if $n\geqslant p$,
one can use the relation $[ \ana^n,\adag^p ]={[ \ana^p,\adag^n ]}^\dagger$.
Such a commutator is computed using basic couting arguments, and one finds 
($M_i=\adag^i \ana^i$)
\begin{equation}
  \label{eq:comm_an_adagp}
  \left[ \ana^n,\adag^p \right]=\adag^{p-n}\sum_{c\geqslant 1} \left(c!C_n^c C_p^c\right) M_{n-c}, 
\end{equation}
$C_n^p=n!/[p!(n-p)!]$ being the usual binomial coefficient for $p\leqslant n$, 
and being 0 otherwise. 
The sum in (\ref{eq:comm_an_adagp}) is performed over the number $c$ of 
contracted $\ana$ and $\adag$ operators.


\subsection{Derivation of the Flow Equations}
\label{app:sub:der_flow_eq}

Here the evaluation of the commutator of the generator with the Hamiltonian is
addressed. Note that for Wegner's generator the first task is to 
compute the generator, given by the commutator of two parts of the Hamiltonian.
As the computation of $\ew$ is qualitatively not different from the 
computation of $[\eta,H]$ we will not discuss it here but focus on the 
MKU-generator. We will however use notations which are more general than the 
ones used in the main text, and that are easily transferable to 
Wegner's generator. 

The flowing Hamiltonian has the form 
\begin{equation}
  \label{eq:ham_general_form_MKU_1}
  H=H_0+\sum_{k=2,4} \left(H_k^++H_k^-\right),
\end{equation}
with
$H_k^-=\left({H_k^+}\right)^\dagger$ and
\begin{equation}
   \label{eq:ham_general_form_MKU_2}
   H_0=\sum_{i\in \mathbb{N}}
   g_{0,i} M_i
   \;\mbox{ and }\;
   H_k^+=\sum_{i\in \mathbb{N}} g_{k,i} \adag^k M_i.
\end{equation}
For Wegner's generator, the sum over $k$ in (\ref{eq:ham_general_form_MKU_1}) 
would extend over all non-negative even integers. The MKU generator simply 
reads 
\begin{equation}
  \emku=\sum_{k=2,4} \left(\eta_k^+-\eta_k^-\right)
  =\sum_{k=2,4} \left(H_k^+-H_k^-\right),
\end{equation}
from which one deduces the flow equations
\begin{subequations}
  \label{eq:flow_eq_very_general}
  \begin{eqnarray}    
    \pal H_0&=&2\left(\left[H_2^+,H_2^-\right]+\left[H_4^+,H_4^-\right]\right)\\
    \pal H_2^+&=&\left[H_2^+,H_0\right]+2\left[H_4^+,H_2^-\right]\\
    \pal H_4^+&=&\left[H_4^+,H_0\right].
  \end{eqnarray}
\end{subequations}

As an example, let us now outline the computation of the commutator 
$[H_p^+,H_n^-]$ for $n\leqslant p$. It is equal to
\begin{subequations}
  \begin{eqnarray}
    &&\sum_{i',i''} g_{p,i'} g_{n,i''} \left[\adag^p M_{i'}, M_{i''}
      \ana^n\right]\\
    &=&\adag^{p-n}\sum_{i',i''} g_{p,i'} g_{n,i''}\sum_{c} 
    \mathcal{A}_{i',i'',c}^{p,n} M_{i'+i''+n-c}\quad\quad\\
    &=&\adag^{p-n}\sum_i M_{i}\nonumber\\
    \label{eq:ex_comm}
    &&\times\left(\sum_{i',c} \mathcal{A}_{i',i-i'-n+c,c}^{p,n} 
      g_{p,i'} g_{n,i-i'-n+c} \right),
  \end{eqnarray}
\end{subequations}
where we defined 
\begin{equation}
  \mathcal{A}_{i',i'',c}^{p,n}=c!\left[ C_{i'}^{c} C_{i''}^{c} 
    - C_{i'+p}^{c} C_{i''+n}^{c} \right].
\end{equation}
In (\ref{eq:ex_comm}), the $i$ runs over $\mathbb{N}$, but the sums 
over $i'$ and $c$ are performed on a finite set of positive numbers, 
because $C_n^p=0$ if $p>n$. 
Note that by setting $n=0$ in (\ref{eq:ex_comm}), one obtains $[H_p^+,H_0]$

Collecting everything gives
\begin{subequations}
  \label{eq:flow_eq_explicit}
  \begin{eqnarray}
    \pal g_{0,i}&=&2\sum_{i',c} \left[ \mathcal{A}_{i',i-i'-2+c,c}^{2,2} 
      g_{2,i'} g_{2,i-i'-2+c}\right.\nonumber\\
    &&\quad\quad\left.+\mathcal{A}_{i',i-i'-4+c,c}^{4,4} 
      g_{4,i'} g_{4,i-i'-4+c}\right]\quad\\
    \pal g_{2,i}&=&\sum_{i',c} \left[ \mathcal{A}_{i',i-i'+c,c}^{2,0} 
      g_{2,i'} g_{0,i-i'+c}\right.\nonumber\\
    &&\quad\quad\left.+2\mathcal{A}_{i',i-i'-2+c,c}^{4,2} 
      g_{4,i'} g_{2,i-i'-2+c}\right]\quad\\
    \pal g_{4,i}&=&\sum_{i',c} \left[ \mathcal{A}_{i',i-i'+c,c}^{4,0} 
      g_{4,i'} g_{0,i-i'+c}\right].
  \end{eqnarray}
\end{subequations}
The flow of the observable $X(l)$ can be computed in the same way. One has 
\begin{equation}
  X=\sum_{k\in 2\mathbb{N}+1} (X_k^+ + X_k^-),
\end{equation}
with
$X_k^-=\left({X_k^+}\right)^\dagger$ and
\begin{equation}
  X_k^+=\sum_{i\in \mathbb{N}} \alpha_{k,i} \adag^k M_i.
\end{equation}
The flow equations are found to be
\begin{subequations}
  \begin{eqnarray}
    \pal X_1^+&=&\left[ H_2^+,X_1^-\right]-\left[ H_2^-,X_3^+\right]\nonumber\\
    &&+\left[ H_4^+,X_3^-\right]-\left[ H_4^-,X_5^+\right]\\
    \pal X_3^+&=&\left[ H_2^+,X_1^+\right]-\left[ H_2^-,X_5^+\right]\nonumber\\
    &&+\left[ H_4^+,X_1^-\right]-\left[ H_4^-,X_7^+\right]\\
    \pal X_{k\geqslant 5}^+&=&\left[ H_2^+,X_{k-2}^+\right]
    -\left[ H_2^-,X_{k+2}^+\right]\nonumber\\
    &&+\left[ H_4^+,X_{k-4}^+\right]-\left[ H_4^-,X_{k+4}^+\right].\quad
  \end{eqnarray}
\end{subequations}
These equations can be written in an explicit form, as we did in 
transforming (\ref{eq:flow_eq_very_general}) into (\ref{eq:flow_eq_explicit}). 
But we refrain from doing so, since they would be more complicated than 
(\ref{eq:flow_eq_explicit}). All relations needed to derive them is contained 
in this Appendix.




\end{document}